# Measuring the Sources of Taste-Based Discrimination Using List Experiments[♠]

Ariel Listo[♥]

Ercio A. Muñoz[♦][*]

Dario Sansone[♣]

March 2025

**Abstract**

This study examines how attitudes among supervisors, co-workers, and customers contribute to discrimination against sexual minorities in the workplace. A large, nationally representative sample in Chile was recruited in collaboration with a local firm. The survey employs a series of double list experiments designed to measure attitudes on sensitive issues while reducing social desirability bias, followed by direct questions on attitudes toward sexual minorities. Findings reveal a discrepancy between reported and actual comfort levels with gay individuals in the labor market. Respondents underreport their discomfort by 15 to 23 percentage points, with the largest bias and lowest comfort levels observed when asked about supervising a gay employee. Additionally, respondents consistently underestimate broader societal support for gay employees and co-workers. These differences are reflected in real-stakes donation behavior: respondents who choose not to donate any amount from a lottery to a local LGBTQ-related NGO also report lower comfort levels and exhibit greater misreporting.

---

[♠] Financial support through the Inter-American Development Bank ESW RG-E1952 is gratefully acknowledged. The views expressed in this paper are those of the authors and should not be attributed to the Inter-American Development Bank. We thank Caridad Araujo, Anne Boring, Keith Marzilli Ericson, Adrian Rubli, and Javiera Selman for their helpful comments. We also thank seminar participants at the AEA LGBTQ+ Virtual Seminar Series, the European Commission (Joint Research Centre), and the University of Oxford for their feedback. All errors are our own.

[♥] University of Maryland. E-mail: alisto@umd.edu
[♦] Inter-American Development Bank. E-mail: erciom@iadb.org
[♣] University of Exeter and IZA. E-mail: d.sansone@exeter.ac.uk
[*] Corresponding author

# 1. Introduction

A large share of studies on the economics of gender, race, disability, sexual orientation and gender identity, as well as other legally protected characteristics, focus on testing whether there is any evidence of discrimination in the labor market. These studies predominantly rely on observational data (Blau and Kahn 2017) – using methodologies such as the Kitagawa-Blinder-Oaxaca decomposition – or on correspondence experiments (Neumark 2018; Lippens, Vermeiren, and Baert 2023) – that is, sending fictitious comparable CVs with varying selected features to real employers with job openings. A much smaller number of studies try to link the patterns of discrimination to a specific theory (Bertrand and Duflo 2017): almost all of these studies aim at comparing statistical discrimination (Arrow 1973) versus taste-based discrimination (Becker 1971) – although other studies have expanded these models or looked at different kinds of discrimination such as institutional or systemic discrimination (Small and Pager 2020; Bohren, Hull, and Imas 2023; Onuchic 2024). An even more limited set of studies attempts to uncover the potential source of taste-based discrimination (Dolado, Minale, and Guerra 2023): that is, whether the discrimination stems from employers', co-workers', or customers' preferences not to interact with members of a particular group.[1] This study relies on a series of double list experiments to advance this literature focused on identifying and measuring the sources of taste-based discrimination in the context of discrimination against sexual minority individuals in the workplace in Chile. Importantly, it exploits a large weighted representative sample – one of the largest ever used to conduct list experiments.

Clearly understanding the drivers of discrimination has important policy implications. For instance, statistical discrimination based on incorrect beliefs may require informational campaigns to correct such beliefs. On the other hand, taste-based discrimination due to employers' preferences should require properly enforced anti-discrimination laws, while such laws are less likely to reduce taste-based discrimination due to customers' preferences. Therefore, to effectively inform policymakers, researchers need to not only uncover evidence of discrimination but also analyze its causes. By doing so, this study contributes to the LGBTQ+ economics literature and, more broadly, to the entire literature on discrimination.

Furthermore, measuring attitudes can be in itself policy relevant: attitudes can affect health and socioeconomic behaviors, outcomes, and disparities (Aksoy, Chadd, and Koh 2023; Glasman and Albarracín 2006; NASEM 2020), as well as occupational sorting (Plug, Webbink, and Martin 2014; Gutiérrez and Rubli 2024b). Attitudes can also directly impact minority individuals through minority stress – that is, stress due to internalized homophobia and transphobia, anticipated rejection, constant efforts to hide one's identity, and actual experiences of discrimination and violence (Meyer, 1995). In addition, while there is evidence of positive impacts of employment anti-discrimination laws (Donohue and Heckman 1991; Klawitter and Flatt 1998; Neumark and Stock 2006; Klawitter 2011; Button 2018; Neumark et al. 2019), the effectiveness of such employment protections depends on compliance and the level of support they receive: if employers have a high distaste for sexual minority individuals (or if they believe that other employees or customers may dislike interacting with sexual minority individuals), they will try to find ways to elude these laws.

---
[1] For instance, (Borm et al. 2020) found suggestive evidence from lab experiment with student participants of co-worker and customer taste-based discrimination against transgender workers, but not for employer taste-based discrimination. Relatedly, (Bar and Zussman 2017) showed that Jewish Israelis are willing to pay a premium to receive services from Jewish rather than Arab worker, while (Kelley et al. 2025) provided evidence of gender-based customer discrimination in a randomized field experiment varying the names of online sale agents.

Relatedly, the stated level of support for certain groups or policies may actually impact voting behavior (Friese et al. 2012; Castanho Silva, Fuks, and Tamaki 2022). Indeed, (Stephens-Davidowitz 2014) argued that indirect measures of local racial animosity were larger than estimates obtained from direct survey questions and correlated with voting results in the US presidential elections In this study, in order to link individuals' responses to actual behavior, respondents participate in a lottery for a chance of winning roughly USD 100 extra: they are then asked to decide how they would allocate this extra funding in case of winning the lottery between themselves and a local NGO promoting LGBTQ+ equality.

While there is substantial literature analyzing discrimination based on gender, age, race, and disability (Goldin 2014; Blau and Kahn 2017; Neumark 2018; Goldin 2021; Lippens, Vermeiren, and Baert 2023), the number of studies focusing on LGBTQ+ discrimination is substantially smaller – although rapidly raising (Sansone 2019; Badgett, Carpenter, and Sansone 2021; Badgett et al. 2024). This growth in LGBTQ+ data and research is matching the rise in the share of individuals identifying as LGBTQ+ (Jackson 2023). Most of the previous studies have found large disparities in the labor market affecting sexual and gender minority people: for instance, wage penalties for gay men and bisexual individuals have been documented in numerous countries, and LGBTQ+ individuals are less likely to be invited for job interviews (Badgett et al. 2024).

Given these stylized facts, there is a clear need to continue documenting the challenges faced by LGBTQ+ individuals. First, previous reports have highlighted that it would be particularly valuable to study the attitudes of supervisors and co-workers of LGBTQ+ people (NASEM 2020). Second, expanding the analysis of socioeconomic disparities by sexual orientation and gender identity to Latin American countries is essential to understand whether these minorities face the same kind of inequalities documented in high-income countries. Without such analyses, these minorities will remain invisible: policymakers will be able to ignore LGBTQ+ individuals as long as disparities and barriers facing these minority individuals are not documented.

One key methodological challenge in measuring attitudes towards sexual minorities among employers, co-workers, and customers is that people may misreport their preferences and beliefs when asked directly in a survey because, for instance, they may fear their answers are not socially acceptable. This is usually referred to as social desirability bias or sensitivity bias (Blair, Coppock, and Moor 2020). These preferences are also hard to detect from actual behavior since employers', co-workers', and customers' choices to hire or interact with sexual minority individuals are influenced by a wide range of factors, such as the number and quality of available alternatives, legal constraints, and beliefs on differential productivity between sexual minority and heterosexual individuals.

To overcome these barriers, this study relies on a survey experiment called a list experiment: individuals are presented with a list of statements and asked to report how many of the statements in the list are true for them (without specifying which statements are true for them). In the list experiments analyzed in this paper, one group of respondents is presented with four non-key statements, and another group is presented with the same four statements plus an additional key statement pertaining to their level of comfort supervising a gay employee, working closely with a gay colleague, or having a gay cashier at the grocery store. Comparisons in the average number of items reported across lists allow researchers to estimate the true share of respondents who agree with each key statement of interest regarding sexual minority individuals. More specifically, this study relies on double list experiments to verify the robustness of the findings to using different non-key statements (Chuang et al. 2021) and to increase the precision of the estimates (Droitcour et al. 1991; Glynn 2013).

While the list experiment technique cannot identify which specific individuals agree with the key statements (because individuals only report the total number of statements within each list that are true for them, as opposed to indicating whether each statement is true for them), it allows researchers to credibly estimate population-level estimates on sensitive issues while removing social desirability bias. One can then estimate the magnitude of such social desirability bias by comparing the estimates from the list experiments to the average responses to questions directly asking individuals about their level of comfort supervising a gay employee, working closely with a gay colleague, and having a gay cashier at the grocery store.

Our results indicate that individuals underreport their discomfort with gay individuals in the labor market by 15 to 23 percentage points. The highest gap arises from supervisors, who report the lowest level of comfort with gay employees. In addition, these differences correlate with real-stakes donation behavior: the degree of overreporting is higher and the level of comfort is lower among individuals who chose not to donate any amount to LGBTQ-related NGOs. It is worth noting that respondents are specifically asked about their level of comfort – not whether they think a gay employee, co-worker, or cashier would be more productive – to focus the analysis on taste-based discrimination rather than on statistical discrimination.

List experiments have been used extensively in other social sciences, such as sociology and political science (Rayburn, Earleywine, and Davison 2003; Lax, Phillips, and Stollwerk 2016; Gervais and Najle 2018; Streb et al. 2008; Blair, Coppock, and Moor 2020; Li and Van den Noortgate 2022), including in Chile to study vote buying (de Jonge 2015), but less so in economics (Coffman, Coffman, and Ericson 2017; Aksoy, Carpenter, and Sansone 2025; Jamison, Karlan, and Raffler 2013; Chuang et al. 2021; Boring and Delfgaauw 2024; McKenzie and Siegel 2013; Agüero and Frisancho 2022).[2] List experiments work particularly well for measuring attitudes – as in this study – rather than behavioral or personal characteristics (Ehler, Wolter, and Junkermann 2021). In addition, as already emphasized in (Osman, Speer, and Weaver 2025), using list experiments to study discrimination in developing countries can be particularly valuable since observational analysis using Kitagawa-Oaxaca-Blinder decomposition are limited by what variables can be directly observed, while correspondence experiments are less feasible in economies with lots of small firms not relying on resumes or job websites, thus underestimating the extent of discrimination, as small firms are the ones more likely to be biased.

The empirical section includes heterogeneity analyses estimating the level of comfort and the social desirability bias among specific subgroups based on characteristics such as sex, education, income, employment status and managerial experience, political affiliation, or religiosity. These estimates indicate higher levels of comfort among women, those with managerial experience, those with left-leaning political affiliations, and non-religious respondents. However, misreporting is widespread across most groups.

These survey experiments are then followed by standard sociodemographic questions and additional opinion questions. These questions allow the comparison between the respondents' level of support with gay individuals and other sexual minority individuals, as well as other minorities, and in different environments. Our findings suggest that the stated level of comfort tends to be lower for interaction with gay men in the workplace compared to other contexts or other minorities.

In addition, respondents are asked to guess the shares of the Chilean adult population who would be comfortable with a gay employee, co-worker, or cashier. These shares can be compared to the estimates from the list experiments to test whether individuals underestimate or overestimate the level of comfort with sexual minority individuals in the general population. Individuals' beliefs about population attitudes toward gay employees and co-workers are consistently lower than their own elicited support, even after accounting for the social desirability bias.

Chile is an ideal context for studying these issues for several reasons. First, with a few exceptions (Muñoz and Sansone 2024; Muñoz, Saavedra, and Sansone 2024a; Muñoz, Sansone, and Ysique 2024; Nettuno 2024; Tampellini 2024; Muñoz, Saavedra, and Sansone 2024b), most LGBTQ+

---

[2] There are two recent list experiments worth highlighting since they have also been conducted in Latin America to estimate the size of the LGBTQ+ population: (Ham, Guarin, and Ruiz 2024) in Bogotá, Colombia, and (Gutiérrez and Rubli 2024a) in Mexico. (Gutiérrez and Rubli 2024a) used a list experiment to also test whether people would rather work with a straight person, but their estimate from the list experiment was counter-intuitive and lower than the average from the direct question: the authors acknowledged that these findings may have been driven by people misinterpreting the question. We conducted extensive piloting and cognitive interviews to avoid encountering the same problem.

studies have focused on high-income countries, especially in North America and Western Europe.[3] Therefore, there is a strong need to reduce the invisibility of LGBTQ+ individuals in low- and middle-income countries. Second, Chile has a developed formal labor market, thus it is more appropriate to study these issues of workplace discrimination (e.g., co-worker preferences).

Third, Chile has made significant legislative advances in LGBTQ+ rights in the past few decades, especially after the ending of the military dictatorship in 1990: same-sex sexual activities were decriminalized in 1999, an employment anti-discrimination law was passed in 2012, and same-sex marriage was legalized in 2021. At the same time, attitudes towards LGBTQ+ individuals are also generally positive: most people support the right of same-sex couples to get married, and think that sexual minority individuals should be permitted to run for public office (Muñoz, Sansone, and Ysique 2024). Similarly, 78% of respondents in our sample know someone who is gay among their immediate family, relatives, neighbors, co-workers or friends (Table B1). However, there is evidence of recent backlashes (Palacios 2024), gender norms remain conservative and gender-based discrimination is still widespread (Montoya et al. 2025), so it is important to understand whether the majority of the Chilean population still supports LGBTQ+ individuals and whether such support is context-specific or limited only to certain sexual and gender minority individuals.

## 2. Data and methodology

### 2.1. List experiments and survey design

The main analysis relies on the list experiment technique (also called "item-count technique", "unmatched count", or "veiled approach") that was pioneered by (Raghavarao and Federer 1979) and (Miller 1984). As mentioned in the introduction, respondents are given a list of statements and are asked to report how many statements (but not which specific ones) are true for them, thus providing an extra layer of anonymity and increasing privacy (Coutts and Jann 2011). Participants are either assigned to a treatment group or a control group. In the control group ("short list"), participants are given a list of statements and asked to indicate how many of those statements are true for them. In the treatment group ("long list"), participants are given the same list of statements plus a key statement of interest (in this study, a statement about their level of comfort with gay individuals in different contexts). The difference in means between the two lists can be interpreted as the estimated share of the population with the key attribute of interest. Table 1 presents one of the lists used in this study (translated from the original Spanish version).

One can then extend this technique by using double list experiments. For each key sensitive statement, participants are presented with two lists (List A and List B) whose items are designed to be positively correlated. Each list contains four non-key statements. Half of the participants (randomly selected) see List A (a short list) and then List B with the key statement (a long list).

---

[3] Another study worth mentioning is (Abbate et al. 2024): the authors conducted a correspondence experiment in the rental housing market in four Latin American countries and found evidence of statistical discrimination against couples with a transgender woman.

The other half see List A with the key statement (a long list) and List B (a short list). The differences in means between short and long lists from both lists A and B are averaged to provide the estimated share of the population with that key sensitive attribute. Thanks to this extension, it is possible to obtain more precise estimates by increasing power and reducing variance (akin to the advantages of a within-subject design) since all respondents provide information about all key statements, unlike the single list experiment in which only respondents seeing the long list provide such information (Droitcour et al. 1991; Glynn 2013). An additional advantage of using the double list method is that it allows to verify the robustness of the main findings to using different non-key statements by comparing the estimates obtained from List A versus List B (Chuang et al. 2021).

This study includes three sets of double list experiments with three key statements:

*Supervisor distaste*: "I would feel comfortable supervising a gay employee."

*Co-worker distaste*: "I would feel comfortable working closely with a gay co-worker."

*Customer distaste*: "I would feel comfortable having a cashier at the supermarket who is gay"

The double list experiment technique is employed for all three statements, thus leading to a total of six lists: Lists 1A and 1B for supervisor distaste, Lists 2A and 2B for co-worker distaste, and Lists 3A and 3B for customer distaste.

Direct questions regarding the key statements are then asked to all participants after they respond to demographic and socioeconomic questions in a survey. The direct questions provide baseline estimates of the share of the population with the key attributes, thus allowing us to estimate the size of the social desirability bias. Ex-ante, the size of this bias is not clear: online surveys may elicit truthful answers since they are self-administered, potentially completed in private, and anonymous (Holbrook and Krosnick 2010; Robertson et al. 2018). Thus, the magnitude of misreporting documented in this study is likely to be a lower bound to what might occur in other surveys, since most surveys are not conducted with as much privacy and anonymity and thus respondents in this study may be less prone to social desirability bias even when answering the question directly.

All subjects first participate in the list experiment section and then advance to the survey. Subjects are not allowed to skip any questions in the list experiments and are not allowed to go back and revise their answers at any point. In addition to the three direct questions (relating to the three key statements from the list experiments) and standard demographic and socioeconomic variables, the survey includes questions on sex, sexual orientation and gender identity, and additional direct questions to measure participants' stated views toward women and minorities in different environments.

Participants are then asked to provide their estimates of the share of the adult Chilean population who would feel comfortable supervising a gay employee, working closely with a gay colleague, or having a gay cashier at the supermarket. Finally, participants are told that they entered into a lottery and are asked to indicate how much they would like to donate from their earnings to a local LGBTQ+ charity if they are among the randomly selected winners (each receiving CLP 100,000, i.e., around USD 100).[4]

## 2.2. Data collection and study sample

The first draft of the questionnaire, list experiments, and pre-analysis plan were reviewed by experts between June and July 2024. The updated experiment protocol was pre-registered in July 2024. After pre-registering the experiment,[5] and obtaining ethical approval from the authors' universities, a pilot was conducted on an online platform (Prolific) using a non-representative sample with 535 respondents from Chile. The findings from this pilot are discussed in Section C of the Online Appendix. Notably, the pilot provided evidence that the list experiments did not affect the answer to the direct questions on attitudes towards sexual minorities (Appendix C.1) and that whether participants expressed comfort or discomfort with gay men were driven by distaste or preferences rather than by productivity beliefs (Appendix C.4).

Following this pilot, a local survey company (Datavoz) conducted 10 cognitive interviews in August 2024. Datavoz then conducted a pilot with 235 participants, followed by a soft launch with 62 participants. The main data were collected by Datavoz from early October until mid-December 2024 and it included 4,000 participants in Chile. The final sample excludes data from the pilot, soft launch and the cognitive interviews and only includes data from the main wave.[6] Datavoz is an established survey company routinely used by other organizations such as the United Nations Development Programme, the International Labour Organization, Vanderbilt University, and the Universidad Católica de Chile.

It is worth emphasizing that the final sample size is larger than in most previous studies. In fact, almost all the list experiments summarized in (Blair, Coppock, and Moor 2020) had fewer than 1,000 respondents. Similarly, only 4 out of the 54 list experiments reviewed in (Li and Van den Noortgate 2022) had more than 1,000 observations in the control group (short list) and 1,000 in the treatment group (long list). This is likely to have been the key factor in some of these studies not having enough power to detect differences in prevalence rates or social desirability bias, especially across subgroups.[7]

---

[4] Section A.1 of the Online Appendix discuss ethics and pre-registration. Section A.2 presents additional technical details on the list experiment technique and the sources of sensitivity bias. Section A.3 discusses the randomization of the items within each list experiment, as well as the randomization of the order of the list experiments: there is little evidence of answers to the list experiments being affected by the order in which participants saw the lists. Section A.4 discusses several list experiment design considerations (e.g., selecting the number and type of non-key items, and avoid priming respondents), provides evidence that flooring and ceiling effects are negligible in these list experiments, and describes additional advantages from using survey experiments and online surveys. Section A.5 provides additional evidence supporting the validity of the list experiment assumptions (treatment randomization, no design effect, and no liar).

[5] See https://www.socialscienceregistry.org/trials/13989.

[6] We provide unweighted results including data from the pilot and the soft launch in Section A.6 of the Online Appendix. The main findings remain qualitatively similar, with a slightly larger level of support estimated by the double list experiment from customers, which renders a slightly lower estimated social desirability bias for that group.

Participants in the main study were recruited based on sex, region, and age quotas with the goal of achieving a representative sample of the Chilean population along these three dimensions. The questionnaire and list experiments were coded on Qualtrics. The survey was tested to work effectively and easily on multiple platforms (Windows, Apple, and Android) and was designed to be mobile-friendly. The recruitment email and survey used by Datavoz are reported in Section E of the Online Appendix. Participants never disclosed any identifying information, and the survey was completely anonymous. In our analyses below, we include all respondents who finished the experiment and the survey in their entirety.[8]

The main questionnaire took about 14 minutes to complete (median response time). Fifty gift cards of around CLP 50,000 (USD 50) each were then raffled among the participants who completed the survey as participation payments. Additionally, up to 50 additional gift cards of roughly CLP 100,000 (USD 100) were raffled among all respondents as part of the donation question.

Table B1 presents weighted summary statistics of the Datavoz participants. Comparing the sample to official population estimates from the Census and other national representative surveys, the main weighted sample appears representative not only based on age, region, education, and sex – as expected given the sampling and weighting methodology – but broadly also with respect to several other variables such as religious affiliation (32% of respondents reporting no religious affiliation, compared to 29% of respondents from 2023 Latinobarometro from Chile), political affiliation (55% of respondents "lean left", compared to 47% of respondents from the 2018 World Value Survey from Chile), and indigenous status (10% reporting indigenous descent, compared to 13% of respondents from the 2017 Chilean Census). Similarly, the weighted average income range is between CLP 1,200,001 and CLP 1,450,000, while the median income range is between CLP 975,000 and CLP 1,2000,000, thus close to the average income of CLP 1,304,771 estimated in the 2022 Encuesta de Caracterización Socioeconómica Nacional (CASEN). Moreover, the estimated share of sexual and gender minorities is consistent with previous online surveys (Jackson 2023).

One may argue that the main drawbacks of these kinds of online samples are that it is difficult to estimate the margin of error for the general population and that they do not include respondents from the non-Internet population. However, as noted in (Haaland, Roth, and Wohlfart 2023), given that most probability-based panels have relatively high nonresponse rates, the differences in the extent of selection between probability-based samples and quota-based online datasets might not be that large in practice. Moreover, (Haaland, Roth, and Wohlfart 2023) summarize evidence from several studies arguing that the online and offline populations hardly differ in terms of survey responses and experimental results. It is also reassuring to note that 90 percent of the Chilean population used an Internet connection in 2021, comparable to 92 percent in the US (World Bank 2022).

---

[7] One exception is the study in Mexico by (Gutiérrez and Rubli 2024a), which had a larger sample size (around 10,000) given their focus on measuring LGBTQ+ identification.

[8] Section A.6 in the Online Appendix reports additional data quality checks – such as checking that participants paid attention, excluding respondents who provided the same answer for all list experiments, excluding respondents who completed the survey too quickly or too slowly, or using unweighted data – and discusses how standard errors for the main analysis were computed.

## 3. Results

This section consists of five parts. First, we present the main results from our double list experiments. Second, we discuss how our estimates from the list experiments compare with population beliefs. Third, we explore heterogeneity analyses based on a set of relevant participant characteristics. Fourth, we compare results on stated attitudes towards gay individuals to attitudes towards other sexual minorities and other minority groups. Lastly, we map the findings from our double list experiment and from the direct questions to behavior on a real-stakes, incentive-compatible donation question.

### 3.1 Main results from the list experiments

Our main results stem from a comparison of our findings from the double list experiment and the direct questions on attitudes from gay supervisors, co-workers, and customers. These weighted results are presented in Figure 1. The first two bars (from left to right) of Figure 1 measure attitudes towards supervising a gay employee based on the answers to the list experiments (dark gray bar labeled "Double List") and the direct question (light gray labeled "Direct Question"). The middle two bars calculate attitudes towards a gay colleague. The last two bars estimate attitudes towards a gay cashier. In each of these three sets of dark and light gray bars, the difference between the estimates from the list experiments and the direct questions is reported above the bars. For each category, the double list experiment share has been obtained by averaging the estimates from List A and List B. Weighted and unweighted estimates from Lists A and B separately are reported in Appendix D.

All three sets of bars indicate that discomfort with gay individuals from supervisors, co-workers, and customers is substantially underreported, by roughly 15 to 23 percentage points. The first two bars in Figure 1 indicate that our estimated share of individuals comfortable supervising a gay employee is 57.3%, whereas the share of individuals reporting being comfortable supervising a gay employee in the direct question is 80.7%, a difference of 23.4 percentage points. Similarly, the middle two bars in Figure 1 indicate that our estimated share of individuals comfortable working closely a gay co-worker is 66.9%, whereas the share of individuals reporting being comfortable working closely with a gay co-worker in the direct question is 82%, a difference of 15.1 percentage points. Lastly, the right-most two bars in Figure 1 indicate that our estimated share of individuals comfortable with a gay cashier is 65.7%, whereas the share of individuals reporting being comfortable with a gay cashier in the direct question is 87.4%, a difference of 21.7 percentage points. The social desirability bias estimated in this context is somewhat higher than the average bias detected in previous list experiments (Blair, Coppock, and Moor 2020; Ehler, Wolter, and Junkermann 2021; Li and Van den Noortgate 2022), but similar to other studies measuring attitudes towards sexism and DEI policies in the workplace (Boring and Delfgaauw 2024).

Our next set of results involves investigating these differences using regression analyses. Specifically, our double list experiment design allows us to estimate the following regression model:

$$y_i = \beta_0 + \beta_1 T_i + \beta_2 X_i + \mu_i \qquad (1)$$

where $T_i$ takes a value of 1 if the list includes the key statement (i.e., long list) and 0 otherwise, and $X_i$ is a vector of control variables which includes demographic controls (i.e., subject's age, sex at birth, sexual orientation, and current region-commune of residence), socioeconomic controls (i.e., subject's education level, employment status, income, current religious affiliation, political affiliation, and beliefs about the general level of comfort among the Chilean population with the three key statements), and a set of additional controls (i.e., whether at least one child less than 18 years of age lives in the subject's household, number of people living in the subject's household, marital status, and day-of-week and week-of-sample indicators which denote the day of the week the respondent started the experiment and the week number since sample collection started).

Model (1) is estimated using ordinary least squares (OLS), separately for each list and for each key statement. Results are reported in Table 2. Columns 1-4 present results from List A and columns 5-8 present results from List B. Panel A displays the estimated share of individuals agreeing with the supervisor distaste key statement, Panel B displays the estimated share of individuals agreeing with the co-worker distaste key statement, and Panel C displays the estimated share of individuals agreeing with the customer distaste key statement. In addition to these estimates, each panel reports the estimated bias calculated from the difference between each coefficient and the estimate obtained from the corresponding direct question.

As Table 2 shows, the estimates of $\hat{\beta}_1$, that is, the estimated size of our sample with the correspondent key attribute, are largely consistent and robust to the inclusion of a battery of controls. Additionally, all estimates are substantially smaller than the corresponding levels elicited from the direct questions. The estimated social desirability bias ranges from 15 to 35 percentage points for supervisor discomfort, 7 to 27 percentage points for co-worker discomfort, and 19 to 26 percentage points for customer discomfort.

**3.2 Population beliefs**

The next set of results focuses on respondents' beliefs about attitudes from the general Chilean population about supervising, working with, and shopping from gay individuals. Figure 2 presents these results. Specifically, respondents were asked to report their estimates in response to the statements: "In the adult Chilean population, I think approximately __ out of every 100 people would feel comfortable [supervising a gay employee] (Panel A) / [working closely with a gay co-worker] (Panel B) / [having a cashier at the supermarket who is gay] (Panel C)." This elicitation of beliefs was designed to map directly into the three supervisor, co-worker, and customer distaste key statements introduced in the list experiments.

The histograms in Panels A and B roughly delineate normal distributions: most respondents believe that population-level comfort with gay employees and co-workers lies between 50-60%. On the other hand, Panel C displays a left-skewed distribution: the median is higher than the mean and a substantial fraction of respondents believe that comfort with a gay cashier lies between 90-100%.

In line with previous findings by (Aksoy, Carpenter, and Sansone 2025), we find that both the mean and median beliefs about comfort among the general population in all three panels are below our estimates from the list experiments of the share of our sample that is comfortable with the corresponding key statements. In the exhibit, this comparison is presented in the box plot below each histogram, where the white vertical line "|" indicates the median; the white "+" symbol indicates the mean; and the black "x" symbol indicates the share estimated from the list experiment. This result indicates that individuals may underestimate the share of comfort with and support for gay workers in the labor market, especially regarding supervising gay employees and having gay co-workers.

### 3.3 Heterogeneity analysis

This section focuses on estimating and presenting results on heterogeneous effects of our key findings. Several demographic, socioeconomic and other observable characteristics may moderate the effects presented so far. Most of these observables are introduced as controls in the regressions models presented in Table 2: i.e., subject's sex at birth, age, sexual orientation, education level, employment status, income, religious affiliation, political affiliation, beliefs about the general population, marital status, locality, etc. Table 3 introduces heterogeneous effects of each of these independent variables using an estimation method designed for double list experiments by Tsai (2019).

Several insights surface from the heterogeneity analyses presented in Table 3. Young and middle-aged adults have higher shares of support towards gay employees and co-workers compared to older groups. Women have higher levels of comfort toward gay co-workers and cashiers compared to men. Compared to sexual minorities, heterosexual respondents have lower levels of comfort towards gay individuals in all three roles, and the difference is statistically significant for the customer statement. Higher income and non-college educated individuals report lower levels of comfort as co-workers and customers of gay men, respectively. On the other hand, respondents with managerial experience report higher levels of support across all three levels. Participants from outside the metropolitan region of Santiago report lower comfort with gay co-workers and those who lean politically left report higher levels of comfort as supervisors. Those who do not identify with a religion are significantly more likely to feel comfortable with gay workers in any of the three roles, compared to those who declare a religious affiliation. Lastly, holding more optimistic beliefs about the general population's attitudes towards gay people is related to higher reported levels of comfort.[9]

### 3.4 Comparison with other attitudes

This section focuses on comparing answers from the direct LGBTQ-sensitive questions to answers from direct questions on attitudes toward gay people in other contexts or attitudes regarding other sensitive topics. Specifically, in addition to the three supervisor, co-workers, and customer distaste direct questions, respondents were surveyed on whether they would feel comfortable:

- having a gay boss / dentist / real estate agent / taxi driver / waiter / neighbor.
- having to work closely with a lesbian co-worker.
- having to work closely with an indigenous co-worker.
- supervising several employees.

Figure 3 presents a summary of our findings. Overall, the majority of respondents report that they would feel comfortable in all the scenarios presented. In fact, 66% of respondents answered "Yes" to all direct questions. This result is in line with Chile's recent history of progressiveness in social issues. However, some variation arises across statements. The highest share of comfort reported (92% of respondents) results from the indigenous co-worker statement. On the other hand, the lowest share of (81% of respondents) is prompted by the supervising a gay employee statement.

Consistently throughout our study, the idea of supervising a gay employee has elicited the lowest share of individuals feeling comfortable regardless of the elicitation method. The statement about gay co-workers elicits more support, but slightly lower than the statement about lesbian co-workers and lower than statements about interactions with gay people in shorter or more distant contexts.

Specifically, shorter interactions with gay individuals, such as checking out at the supermarket with a gay cashier, having a gay realtor, having a gay taxi driver or having a gay waiter, results in very high levels of individuals reporting feeling comfortable. Distant interactions, such as having a gay neighbor, as opposed to "closely" working with a gay person, also elicits a high share of individual expressing comfort. These findings suggest that the level of comfort tends to decrease as the perceived intimacy or frequency of interactions increases. While interactions in casual service roles often result in high acceptance, situations that involve sustained or direct collaboration, such as working closely or supervising, appear to evoke more discomfort. This highlights complex dynamics in attitudes, where proximity and role expectations seem to influence reported comfort levels. However, it is important to emphasize that Figure 3 reports stated support: the findings may be different when correcting for social desirability bias. Indeed, level of comfort for gay cashier and gay co-worker is closer when looking at the estimates from the list experiments in Figure 1, thus highlighting that the relationship between comfort level and length of the interaction may be more nuanced.

---

[9] Appendix D reports univariate heterogeneity analyses, as well as additional multivariate analyses with additional controls.

## 3.5 Donation behavior

So far, we have presented results from our double list experiment and survey measures collected from an extensive, nationally representative sample which is considerably larger than samples used in most list experiments in the literature. Next, we investigate the consistency of these results after splitting our sample based on participants' choices in a real-stakes, incentive compatible donation question.

Towards the end of the survey, all participants were entered into a raffle for an additional payment of approximately 100 USD. Respondents were informed that the probability of winning this award was uniform for all participants who finished our study and that it was not affected by their behavior. Following the introduction of the lottery, but before the announcement of the results, individuals were asked whether they would donate any amount of their earnings to one of two local LGBTQ-related NGOs, if they were to be randomly selected to win this prize. Respondents were randomly assigned one of the two NGOs and were provided with their mission statement so as to standardize the minimum amount of information all participants had on the organizations. Participants were informed that winners would have their wishes honored. That is, any amount they chose not to donate was theirs to keep (added to their compensation for participation, if any), and any amount they chose to donate would be sent to the corresponding NGO by our partner, Datavoz, after completion of the experiment.

We found that 1,700 (41.77% of the sample) chose not to donate any of their earnings if they were to win the survey. On the other hand, 2,370 (58.23% of the sample) voluntarily chose to donate some or all their earnings.

## 4. Conclusion

This study provides evidence of distaste towards gay individuals in the workplace. Many respondents in Chile tend to express comfort when asked directly about their views on supervising, working closely, and buying from gay individuals. However, estimates from the double list experiments indicate that many of these respondents are misreporting their attitudes.

These results provide evidence of distaste which could drive taste-based discrimination towards sexual minority individuals in the workplace. Even if the level of discomfort and the estimated social desirability bias are the highest when respondents consider supervising a gay employee, the findings in the other list experiments suggest that distaste is prevalent also among co-workers and customers.

In line with group identity and contact theory, the level of support is the highest among LGBTQ+ respondents and those who know someone who is gay. Women, young people, employed and educated individuals are also more supportive. Similarly, higher comfort levels are found among respondents with managerial experience. Despite these differences and the large estimated social desirability bias, it is important to emphasize that the majority of respondents in all categories express comfort in interacting with gay individuals in the workplace.

When comparing these attitudes to those towards sexual minorities in other contexts or towards other minorities, the evidence suggests that individuals report a higher level of acceptance for indigenous people in the workplace than for sexual minorities. Reported levels of support for gay and lesbian individuals in the workplace are comparable, and the level of support for a gay cashier is similar to those reported in other situations.

In addition, results from the survey show that respondents underestimate the level of support towards gay employees and co-workers among the Chilean population. Lastly, these differences correlate with real-stakes donation behavior: the degree of misreporting is higher and the level of comfort is lower among individuals who chose not to donate any amount to the LGBTQ-related NGOs.

This experiment focused on attitudes towards sexual minority individuals. With few exceptions (Aksoy, Carpenter, and Sansone 2025), we know very little about attitudes towards gender minority individuals such as transgender and non-binary individuals. Future research could also explore how the reported attitudes toward gay individuals in the workplace compare to attitudes in other contexts such as sports, media and entertainment, or education settings. Furthermore, as noted in (Blair, Coppock, and Moor 2020), list experiments address bias in measures of explicit attitudes: more research is needed to measure implicit attitudes towards sexual and gender minorities.

From a policy perspective, these findings highlight that, even if people may report support for minorities because they want to conform to laws and norms, their actual views and behaviors may not be consistent with their stated beliefs. This mismatch may therefore lead to no reduction in the large socioeconomic and health inequalities by sexual orientation documented in the literature even while opinion surveys report increasing acceptance of LGBTQ+ individuals. Cultural changes and actual reductions in conscious and unconscious biases may be needed to generate real and sustained decline in socioeconomic disparities.

**Figure 1: List Experiments on Attitudes Toward Gay Individuals by Supervisors, Co-workers, and Customers**

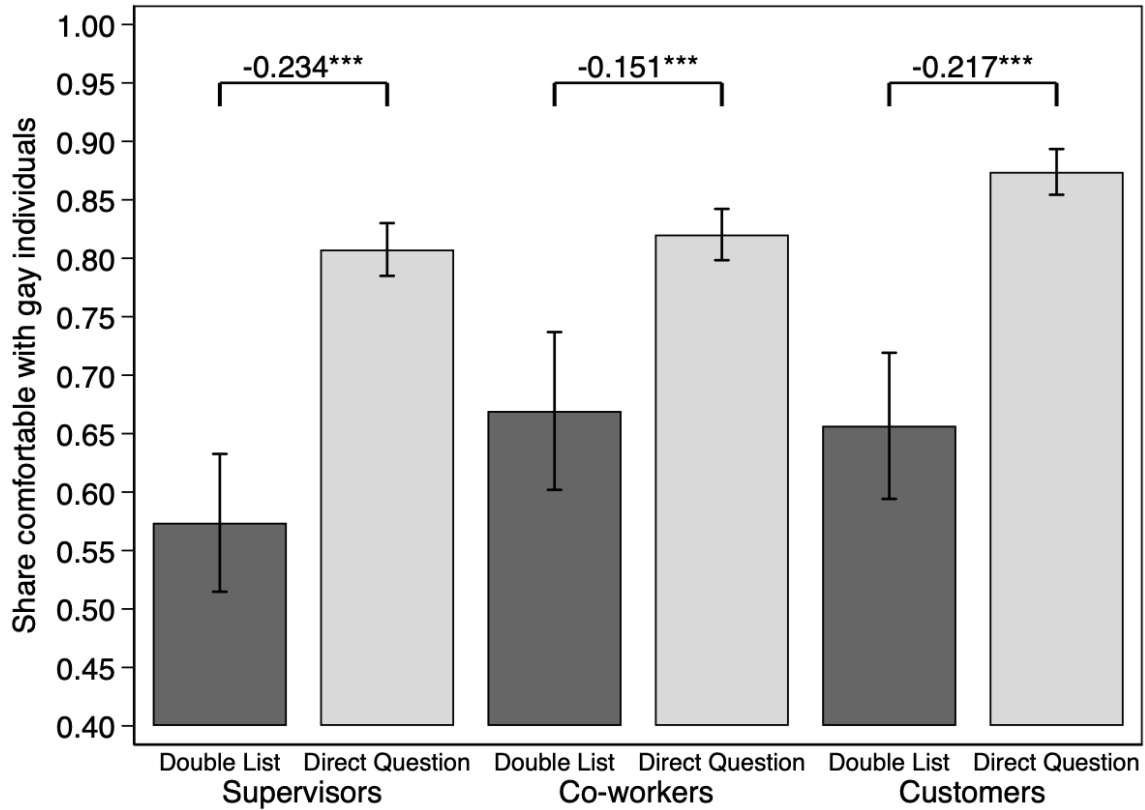

*Notes*: Weighted statistics. 95% confidence intervals are reported with the vertical range plots. The numbers above the horizontal bars are the differences between the two groups at the base of each horizontal bar. Supervisor key statement: "I would feel comfortable supervising a gay employee." Co-worker key statement: "I would feel comfortable working closely with a gay co-worker." Customer key statement: "I would feel comfortable having a cashier at the supermarket who is gay." Number of observations: 4,000. *$p < 0.10$; **$p < 0.05$; ***$p < 0.01$.

**Figure 2. Perceptions of General Views on Attitudes Toward Gay People**

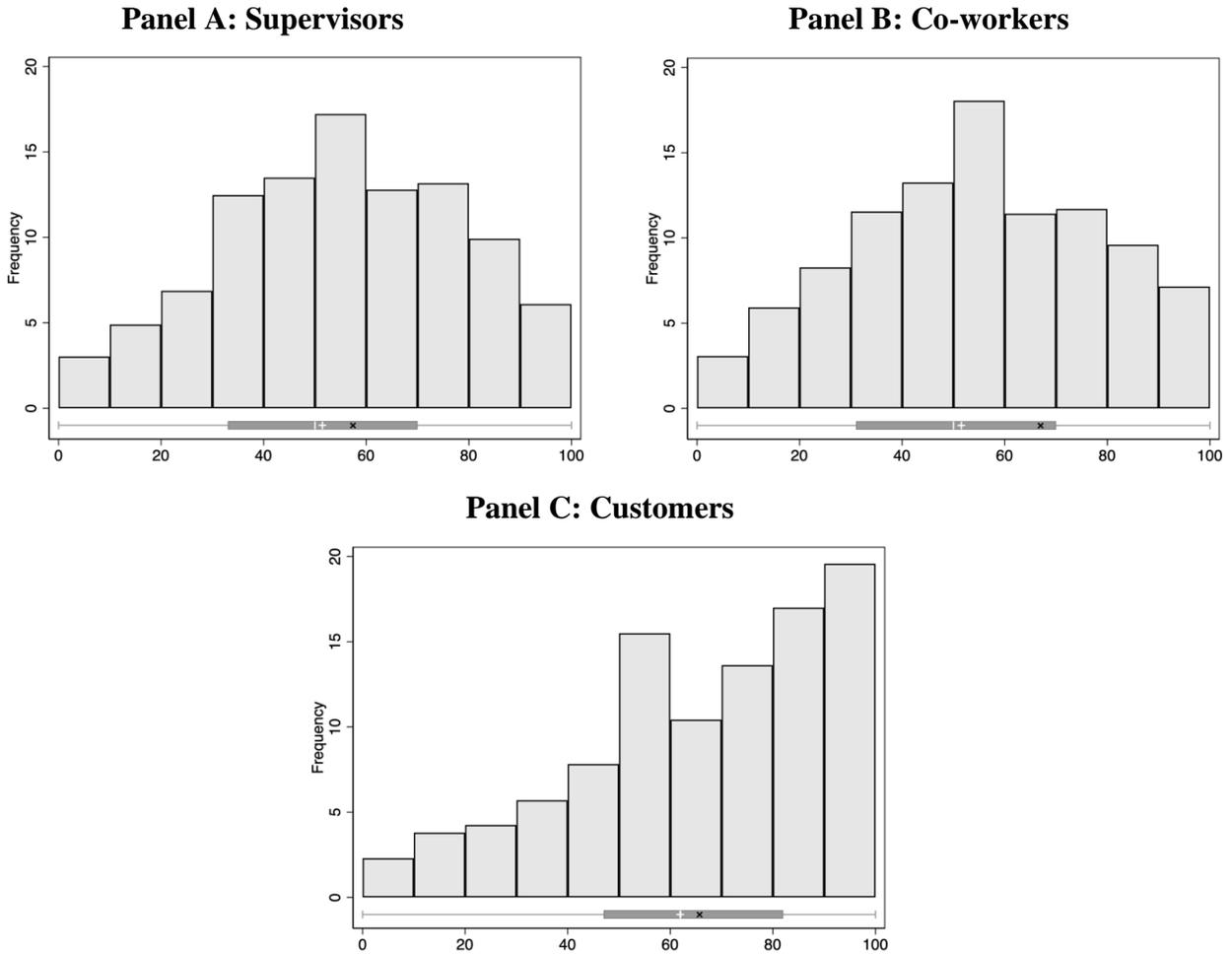

*Notes:* Weighted statistics (and unweighted histogram). The original survey question for panel (a) is "In the adult Chilean population, I think approximately __ out of every 100 people would feel comfortable supervising a gay employee." The original survey question for panel (b) is "In the adult Chilean population, I think approximately __ out of every 100 people would feel comfortable working closely with a gay co-worker". The original survey question for panel (c) is "In the adult Chilean population, I think approximately __ out of every 100 people would feel comfortable having a cashier at the supermarket who is gay." The box plot below each histogram reports minimum and maximum values and 25th and 75th percentiles, as well as mean and median. Within each box plot, the white vertical line "|" indicates the median; the white "+" symbol indicates the mean. The black "x" symbol in panel (a) indicates the actual share of the sample being comfortable supervising a gay employee estimated from the double list experiment in Figure 1. The black "x" symbol in panel (b) indicates the actual share of the sample being comfortable with a gay co-worker estimated from the double list experiment. The black "x" symbol in panel (c) indicates the actual share of the sample being comfortable with a gay cashier. Number of observations: 4,000.

**Figure 3. Comparison of Views Toward Gay Individuals in Multiple Contexts and Toward Other Minority Groups**

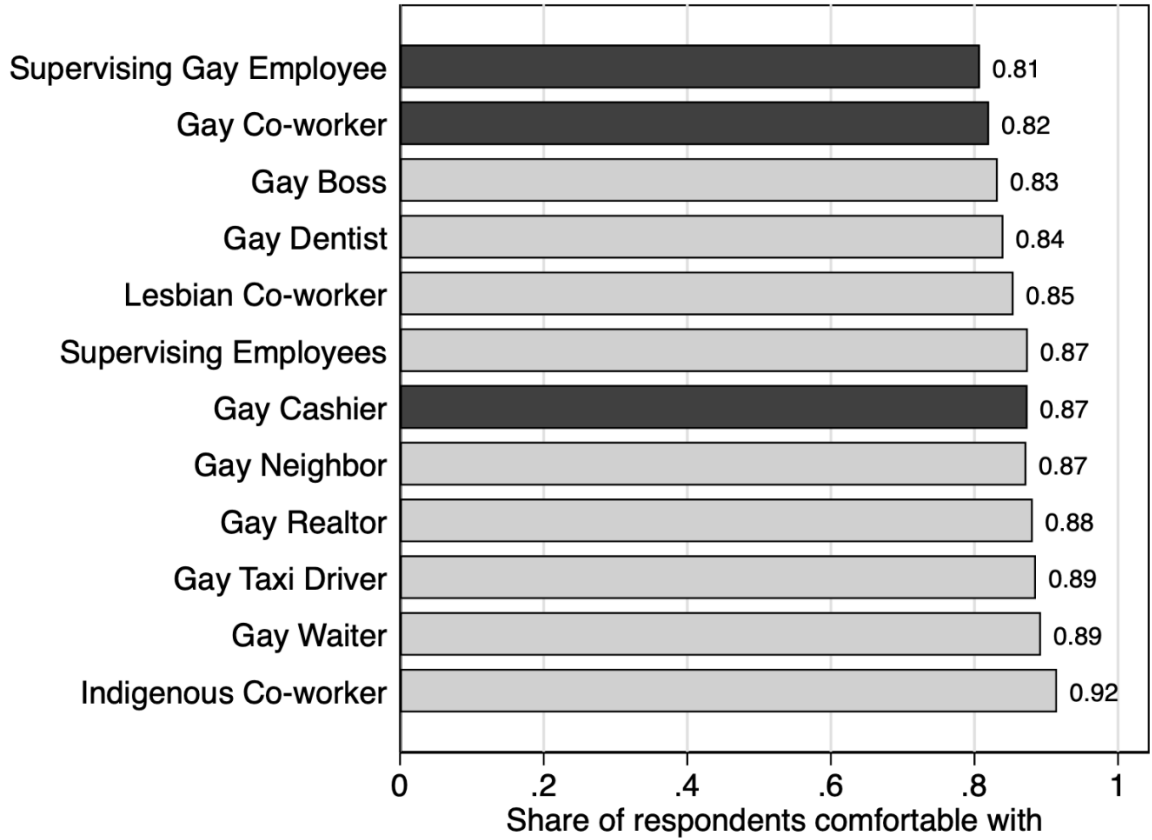

*Notes:* Weighted statistics. Bars in darker gray represent the responses to the three direct questions corresponding to the employee, co-worker, and customer distaste key statements introduced in the list experiments (see Figure 1). The lighter gray bars correspond to the following direct questions (listed in the order they appear in the figure): "Would you feel comfortable… [having a gay boss] / [having a gay dentist] / [having to work closely with a lesbian co-worker] / [supervising several employees] / [having a gay neighbor] / [having a gay real estate agent] / [having a gay taxi driver] / [being served by a gay waiter] / [having to work closely with an indigenous co-worker]?". Number of observations: 4,000.

# Figure 4: List Experiments on Attitudes Toward Gay Individuals by Supervisors, Co-workers, and Customers by Donor Status

**Panel A: Non-donors**

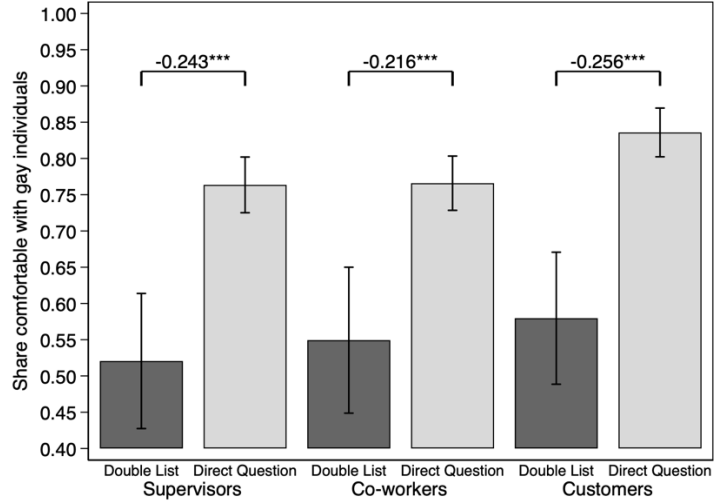

**Panel B: Donors**

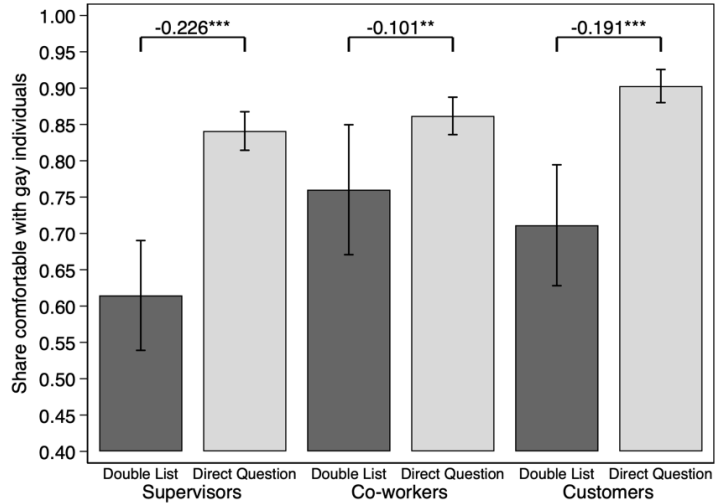

*Notes*: Weighted statistics. 95% confidence intervals are reported with the vertical range plots. The numbers above the horizontal bars are the differences between the two groups at the base of each horizontal bar. See also notes in Figure 1. Panel A: Sample restricted to individuals who chose to not donate any amount of a raffled endowment to one of two local LGBTQ-related NGOs; Number of observations: 1,683. Panel B: Sample restricted to individuals who chose to donate a positive amount of a raffled endowment to one of two local LGBTQ-related NGOs; Number of observations: 2,317. Total number of observations: 4,000. *$p < 0.10$; **$p < 0.05$; ***$p < 0.01$.

**Table 1: List Experiment Example**

| Short list | Long list |
|---|---|
| I own a car or motorcycle. | I own a car or motorcycle. |
| I have a lot of confidence in political parties. | I have a lot of confidence in political parties. |
| I think the military should work with the police to fight crime. | I think the military should work with the police to fight crime. |
| I believe that half of the legislators in Congress should be women. | I believe that half of the legislators in Congress should be women. |
| | I would feel comfortable having a cashier at the supermarket who is gay. *[key statement]* |

*Notes*: The order of the statements within each list was randomized at the subject level. For the full set of lists, as well as the original Spanish questionnaire, see Section E of the Online Appendix.

**Table 2: List Experiments on Attitudes Toward Gay Individuals by Supervisors, Co-workers, and Customers**

|  | List A | | | | List B | | | |
|---|---|---|---|---|---|---|---|---|
|  | (1) | (2) | (3) | (4) | (5) | (6) | (7) | (8) |
| **Panel A: Supervisors** | | | | | | | | |
| Subject saw list with key statement | 0.676 | 0.664 | 0.647 | 0.678 | 0.472 | 0.480 | 0.491 | 0.505 |
|  | (0.054) | (0.044) | (0.042) | (0.043) | (0.050) | (0.047) | (0.049) | (0.048) |
| $R^2$ | 0.152 | 0.234 | 0.220 | 0.245 | 0.076 | 0.133 | 0.153 | 0.205 |
| Estimated bias | 15.075 | 16.275 | 17.975 | 14.875 | 35.475 | 34.675 | 33.575 | 32.175 |
| **Panel B: Co-workers** | | | | | | | | |
| Subject saw list with key statement | 0.772 | 0.754 | 0.739 | 0.771 | 0.567 | 0.580 | 0.594 | 0.612 |
|  | (0.054) | (0.048) | (0.047) | (0.048) | (0.046) | (0.045) | (0.046) | (0.046) |
| $R^2$ | 0.174 | 0.228 | 0.248 | 0.283 | 0.115 | 0.162 | 0.194 | 0.240 |
| Estimated bias | 6.575 | 8.375 | 9.875 | 6.675 | 27.075 | 25.775 | 24.375 | 22.575 |
| **Panel C: Customers** | | | | | | | | |
| Subject saw list with key statement | 0.685 | 0.684 | 0.670 | 0.706 | 0.628 | 0.637 | 0.648 | 0.666 |
|  | (0.054) | (0.047) | (0.046) | (0.045) | (0.055) | (0.050) | (0.052) | (0.050) |
| $R^2$ | 0.140 | 0.202 | 0.204 | 0.266 | 0.107 | 0.192 | 0.215 | 0.264 |
| Estimated bias | 20.75 | 20.85 | 22.25 | 18.65 | 26.45 | 25.55 | 24.45 | 22.65 |
| **Controls for:** | | | | | | | | |
| Region-commune fixed effects |  | ✓ | ✓ | ✓ |  | ✓ | ✓ | ✓ |
| Demographic controls |  | ✓ | ✓ | ✓ |  | ✓ | ✓ | ✓ |
| Socioeconomic factors |  |  | ✓ | ✓ |  |  | ✓ | ✓ |
| Additional controls |  |  |  | ✓ |  |  |  | ✓ |
| Observations | 4,000 | 3,947 | 3,642 | 3,404 | 4,000 | 3,947 | 3,642 | 3,404 |

*Notes*: Multivariate weighted analysis. OLS estimates with robust standard errors in parentheses. All estimates are highly significant at the 1% level. "Estimated bias" reports the differences (in percentage points) between the estimated percentage of participants who agreed to the key statement in each corresponding column and the estimate obtained from the corresponding direct question (as reported in Figure 1). Demographic controls include subjects' age, sex at birth, sexual orientation, gender identity, and indicators for ethnicity (i.e., African descent) and indigenous status. Region-commune fixed effects include current region of residence or commune for those living in the Metropolitan region of Santiago. Socioeconomic factors include subjects' education level, income, employment status, an indicator for managerial experience, current religious affiliation, and political affiliation. Additional controls include whether at least one child less than 18 years of age lives in the subject's household, number of people living in the subject's household, marital status, LGBTQ+ comfort (as defined in Appendix B), an indicator for whether the participant knows a gay person (as defined in Appendix B), and day-of-week and week-of-sample indicators.

**Table 3: List Experiment on Attitudes Toward Gay Individuals by Supervisors, Co-workers and Customers – Heterogeneity Analyses**

| Interaction of treatment variable with: | Supervisors (1) | Co-workers (2) | Customers (3) |
|---|---|---|---|
| Age: 18–44 | 0.127* | 0.129* | -0.040 |
|  | (0.066) | (0.069) | (0.064) |
| Race: African Descent | -0.335 | 0.008 | 0.084 |
|  | (0.222) | (0.176) | (0.210) |
| Indigenous | 0.021 | 0.141 | -0.032 |
|  | (0.118) | (0.122) | (0.124) |
| Sex assigned at birth: Female | 0.002 | 0.158** | 0.180*** |
|  | (0.063) | (0.068) | (0.065) |
| Sexual orientation: Heterosexual | -0.165 | -0.253 | -0.366*** |
|  | (0.123) | (0.177) | (0.111) |
| Household income: More than $975 | 0.069 | -0.165** | -0.003 |
|  | (0.062) | (0.065) | (0.065) |
| Education: Less than high school | -0.043 | -0.069 | -0.137* |
|  | (0.082) | (0.087) | (0.080) |
| Employment status: Employed | 0.098 | 0.130* | -0.031 |
|  | (0.079) | (0.076) | (0.082) |
| Management Experience | 0.134* | 0.138* | 0.173*** |
|  | (0.072) | (0.077) | (0.067) |
| Region: Outside metro | 0.003 | -0.116* | -0.060 |
|  | (0.058) | (0.063) | (0.059) |
| Political affiliation: Lean left | 0.117* | 0.027 | 0.021 |
|  | (0.066) | (0.074) | (0.067) |
| Current religious affiliation: Not religious | 0.137** | 0.116 | 0.233*** |
|  | (0.069) | (0.076) | (0.072) |
| Belief: 50% or more comfortable supervising gay employees | 0.110* |  |  |
|  | (0.063) |  |  |
| Belief: 50% or more comfortable with gay co-workers |  | 0.170** |  |
|  |  | (0.067) |  |
| Belief: 50% or more comfortable with gay cashiers |  |  | 0.045 |
|  |  |  | (0.070) |
| Constant | 0.326** | 0.580*** | 0.781*** |
|  | (0.164) | (0.213) | (0.164) |
| Observations | 3,646 | 3,646 | 3,646 |

*Notes*: Heterogeneity analysis. Multivariate analysis. Robust standard errors are in parentheses. Coefficients obtained using the Stata command kict ls (Tsai 2019) performing least squares estimation for a double list experiment. The dependent variables are the reported number of true statements for the supervisor of a gay employee lists (Column 1), the gay co-workers lists (Column 2), and the gay cashier lists (Column 3). The treatment variable is an indicator equal to 1 for the first long list (List A) containing the corresponding key statement and the second short list (List B), and equal to 0 for the first short list (List A) and the second long list (List B). *$p < 0.10$; **$p < 0.05$; ***$p < 0.01$.

# Online Appendix for "Measuring the Sources of Taste Discrimination Using List Experiments" (NOT MEANT FOR PUBLICATION)

**Appendix A. Technical details.**

**A.1 Ethics and pre-registration.**

At the beginning of the experiment, respondents signed a consent form (reported in Appendix E.2). Only individuals older than 18 were allowed to take part in this study. This research was approved as exempt by the IRB at Maryland University (#2186752) and was approved by the Research Ethics Committee at the University of Exeter (#6475933).

The experiment and a pre-analysis plan were pre-registered on the American Economic Association's registry for randomized control trials on July 9, 2024, and published online on July 16, 2024 (AEARCTR-0013989): https://doi.org/10.1257/rct.13989-1.0.

**A.2 Formal details on the list experiment technique.**

**A.2.1 Mean comparisons in list experiments.**

The main list experiment analysis follows the standard estimation technique implemented in previous studies (Tsai 2019). Suppose there is a sample of *n* participants. Let $T_i$ be the indicator variable equal to one if participant *i* sees the long list with the key sensitive item instead of the short list, and 0 otherwise. Let $S_i$ be the potential answer to the key statement by participant *i*, and let $R_{i,j}$ be the potential answer to the jth non-key statement by participant *i* (where j=4 in this application). Using the list in Table 1, $S_i = 1$ if participant *i* is comfortable having a gay cashier at the supermarket, and 0 otherwise. Similarly, for example, $R_{i,2} = 1$ if participant *i* thinks that the military should work with the police to fight crime, and 0 otherwise. It is worth remembering that researchers do not observe $S_i$ or $R_{i,j}$. Instead, they observe the total number of statements that are true for participant *i*: $Y_i = T_i S_i + R_i$ where $R_i = \sum_{j=1}^{4} R_{i,j}$. Under certain assumptions discussed in Appendix A.5, the following difference-in-means estimator gives the estimated share of the population with the key attribute (i.e., $E(S_i)$).

$$E(S_i) = \frac{\sum_{i=1}^{n} Y_i T_i}{\sum_{i=1}^{n} T_i} - \frac{\sum_{i=1}^{n} Y_i (1 - T_i)}{\sum_{i=1}^{n} (1 - T_i)} \qquad (1)$$

**A.2.2 Mean comparisons in double list experiments.**

To formally introduce the double list experiment technique, let $Y_i^A$ and $Y_i^B$ be the total number of items in List A and B, respectively, that are true for participant *i*. The estimated share of the population with the key attribute is given by $E^{DL}(S_i)$.

$$E^{DL}(S_i) = \left[ \left\{ \frac{\sum_{i=1}^{n} Y_i^A T_i}{\sum_{i=1}^{n} T_i} - \frac{\sum_{i=1}^{n} Y_i^A (1 - T_i)}{\sum_{i=1}^{n} (1 - T_i)} \right\} + \left\{ \frac{\sum_{i=1}^{n} Y_i^B (1 - T_i)}{\sum_{i=1}^{n} (1 - T_i)} - \frac{\sum_{i=1}^{n} Y_i^B T_i}{\sum_{i=1}^{n} T_i} \right\} \right] / 2 \qquad (2)$$

**A.2.3 Sources of sensitivity bias.**

Social desirability bias, or sensitivity bias, is formalized by (Blair, Coppock, and Moor 2020). This kind of bias can be seen as a form of measurement error, as the response provided by an individual when asked directly about a certain issue or topic is different from the latent true value. Measurement errors can occur for a variety of reasons, such as technical issues, miscommunications between the respondents and the enumerators, or memory recall mistakes. An additional source of measurement error is generated by the sensitivity of the question. Indeed, respondents may misreport their true opinions or beliefs to avoid embarrassment, to project a favorable image of themselves to others (as well as to maintain a good self-image), for fear that their responses may be disclosed to authorities, or as a reaction to questions on topics being considered taboo.

Sensitivity bias can therefore occur if and only if four elements are present. First, a social referent the respondent has in mind when considering how to respond to a survey question (a social referent could be the respondents themselves). Second, a respondent's perception that the social referent can infer the subject's response to the sensitive question either exactly or approximately. Third, a respondent's perception about what response (or nonresponse) the social referent prefers. Fourth, a respondent perception that failing to provide the response preferred by the social referent would entail costs to themself, other individuals, or groups. Costs may be social (e.g., embarrassment), monetary (e.g., fines), or physical (e.g., jail time or personal violence).

These conditions are likely to be met in this specific study as individuals may be embarrassed to disclose to the researchers (and to themselves) that they would dislike interacting with gay individuals as employers, co-workers, or customers. The list experiments can address these sources of sensitivity bias by adding an extra level of privacy through the non-key items, thus removing the bias by addressing the second aforementioned element, that is, making it impossible for the researchers to infer the subject's response. Nevertheless, it is worth mentioning that individuals may still be resistant to answering any questions considered too intrusive or taboo. For this reason, the selected key and non-key items in the main experiments are considered controversial, but not so offensive or shocking that respondents would not consider answering the questions.

**A.3 List randomization and list ordering.**

Both the order of the lists and the order of the items within each list were randomized at the subject level. The order of the statements was randomized at the individual level in both the short and long lists. This served two goals. First, if the order of the items were not randomized and the key statements were listed as last, as done by many papers in this literature, one may worry that seeing a gay-related statement as last in three lists could draw extra attention to the key statements. Second, the order of the statements might also have an impact on subjects' answers. By randomizing the order, one can eliminate any aggregate effect coming from the ordering of the statements.

In addition, although it is common practice in the literature not to randomize the order of the lists, this survey incorporated some randomization into the design to control for potential order effects, thus following the approach in (Aksoy, Carpenter, and Sansone 2025). More specifically, participants were randomly allocated to one of the following six paths:

Path 1: List 1A + KS1, List 1B, List 2A + KS2, List 2B, List 3A + KS, List 3B
Path 2: List 1B + KS1, List 1A, List 2B + KS2, List 2A, List 3B + KS3, List 3A
Path 3: List 2A + KS2, List 2B, List 3A + KS3, List 3B, List 1A + KS1, List 1B
Path 4: List 2B + KS3, List 2A, List 3B + KS3, List 3A, List 1B + KS1, List 1A
Path 5: List 3A + KS3, List 3B, List 1A + KS1, List 1B, List 2A + KS2, List 2B
Path 6: List 3B + KS3, List 3A, List 1B + KS1, List 1A, List 2B + KS2, List 2A

Where KS 1, KS 2, and KS3 stand for the following key statements, respectively:

"I would feel comfortable supervising a gay employee."
"I would feel comfortable working closely with a gay co-worker."
"I would feel comfortable having a cashier at the supermarket who is gay."

Qualtrics's *Evenly Present Elements* feature was used to randomize participants into each path.

Lists 1A through List 3B can be seen in the instructions in Online Appendix E. As can be noted above, half of the participants saw List As first, and the other half saw List Bs first. When the distributions of answers were compared across these orders using a linear combination test of weighted means (e.g., comparing responses to each list in Paths 1-2 to Paths 3-4 or Paths 5-6 – 36 possible combinations), 9 lists presented marginally statistically significant differences, which were all minimal in magnitude (i.e., differences in means of less than 0.2 points). Additionally, when the distributions of answers were compared using unweighted chi-square tests, only 5 out of 36 possible lists presented marginally statistically significant differences.

It is also worth noting that, while the order of the lists was randomized as just described, the order of the questions in the survey section after the list experiments was the same for all respondents.

**A.4 List design.**

**A.4.1 Non-key items in the lists were selected to avoid ceiling and flooring effects.**

While designing the list experiments and choosing the non-key statements, this study followed best practices in the literature (Glynn 2013). For example, it is advised to carefully determine how many non-key statements to include. The number of non-key statements should be neither too low nor too low. The number of key statements should be high enough to avoid a ceiling effect, i.e., participants reporting that all statements are true for them, thus removing the privacy protection provided by the list experiment. At the same time, the number of key statements cannot be too high otherwise respondents may not be able to remember or focus on all statements in the list, thus leading to higher variance and measurement error.

After carefully examining previous studies and noting that (Tsuchiya, Hirai, and Ono 2007) found little impact on list experiment performance when varying the number of non-key items between two and five, it was decided on four non-key statements, as in (Aksoy, Carpenter, and Sansone 2025). To avoid a ceiling effect, a statement expected to be false for most people was included in each list. In addition, each list includes a statement expected to be true for most people to avoid a floor effect, i.e., participants reporting zero items, thus also removing the privacy protection provided by the list experiment.

The remaining two non-key statements were chosen such that they were expected to be negatively correlated: that is, one statement that was likely to be supported by more politically conservative people and another one that was likely to be supported by more politically progressive people. For instance, as shown in Appendix E, the statement "I believe that women should be responsible for the care of children" was expected to apply for conservative respondents, while progressive respondents were expected to support the statement "I believe it is wrong to apply the death penalty, no matter the crime".

Figures A1-A3 can be used to check for ceiling and floor effects. As can be seen in these figures, only a very small share of participants reports the highest and lowest possible items in each of the lists. Thus, it is possible to conclude that the flooring and ceiling effects are negligible in these experiments.

Additionally, if the distributions of responses had followed a uniform distribution, then it would have indicated that most respondents provided random answers (Coffman, Coffman, and Ericson 2017). It is therefore reassuring to observe that the distributions of responses do not follow such a uniform distribution, as shown in Figures A1-A3.

**A.4.2 Negatively and positively correlated non-key items within and between lists.**

The choice of including negatively correlated items in each list has the additional advantage of decreasing variance and increasing power. High variance is often an issue because the key statement is aggregated with a number of non-key statements. To some extent, the additional variance is the cost of the higher perceived privacy protection (Glynn 2013). Therefore, list randomization often produces results that are too high in variance to be statistically significant, especially if the attribute, view, or behavior of interest has low prevalence (Karlan and Zinman 2012).

(Osman, Speer, and Weaver 2025) further discuss large standard errors in list experiments and note that because the list randomization method is based on the difference across two variables, the variance of this difference will be mechanically greater than the variance of a direct question (which is only based on one variable). Indeed, in absence of any sensitivity bias, the standard error on the direct question estimate of the proportion of respondents not feeling comfortable with gay individuals would be the following:

$$\sqrt{\frac{p(1-p)}{N}}$$

where p is the proportion answering yes and N is the sample size.

Helding the sample size constant and assuming that the proportion of people answering yes to the *k* non-key questions was also p and that the answers were independent, the standard error on the list randomization would be the following:

$$\sqrt{\frac{(2k+1)p(1-p)}{N}}$$

This is why the standard error from the list randomization is mechanically larger than the direct question. The variance expressions for both the direct question and the list experiment estimator in the presence of sensitivity bias and/or unbalance design are instead derived in (Blair, Coppock, and Moor 2020).

In order to increase power further in the double list, the non-key statements in Lists A and B were also designed to be positively correlated across lists (Glynn 2013). For example, the statement "I believe that women should be responsible for the care of children" in List 1A was chosen to be positively correlated with the statement "I believe that the poor make little effort to get out of poverty" in List 1B (as reported in Appendix E).

**A.4.3 Sensitive versus non-sensitive non-key items.**

Following (Chuang et al. 2021), in order to draw less attention to the key statements and increase the validity of the list experiment, some of the non-key statements in the lists were political or sensitive in nature. For example, as reported in Appendix E, the lists included items about gender norms, immigration (refugees), death penalty, abortion, taxes, drugs, environmental protection, law enforcement, social protests, and sex education in schools.

Additionally, following (Berinsky 2004), participants were not provided a "don't know" option in the direct question since individuals who held socially stigmatized opinions may have hidden their opinions behind a "don't know" response.

Finally, (Coffman, Coffman, and Ericson 2017) showed that list experiments work better when the stigmatized answer in the related direct question is a "no" instead of a "yes". Thus, the key items and related direct questions were designed such that the socially stigmatized answer was always a "no".

**A.4.4 List example provided to participants.**

To ensure that respondents understood the task, they were presented with an example of a list experiment question and guided through the reasoning behind a certain numerical answer. They were reminded that the question only asked them to report the number of items that were true for them, not which one.

All respondents were then given the option to review the instructions one more time before answering the six list experiment questions.

**A.4.5 Priming.**

Both the recruitment messages (reported in Appendix E.1) and the consent forms (reported in Appendix E.2) stated that the goal of the survey was to understand individuals views and preferences. The description of the study did not specifically mention LGBTQ+ issues in order not to prime respondents or obtain a self-selected sample. 207 participants (5.17% of our sample) clicked on the link to read the entire consent form.

**A.4.6 Placebo tests in previous literature.**

It is important to highlight that increased reporting under the veil of the list experiment is not simply mechanical. Indeed, previous research has shown that list experiments provide increased estimates of prevalence only for stigmatized views: there is no evidence of this technique leading to an increase in reporting of innocuous behaviors (Tsuchiya, Hirai, and Ono 2007; Coffman, Coffman, and Ericson 2017). For instance, (Coffman, Coffman, and Ericson 2017) did not find any significant misreporting when the additional key statement in the longer list was "It has rained once where I live in the last four days."

**A.4.7 Advantages of using survey experiments and online surveys.**

As emphasized in (Stantcheva 2023), surveys are not only a way of collecting data, but they allow researchers to create their own identifying and controlled variation, thus providing a high level of control on the data generating process. In addition, while one can recognize that administrative data are great resources, it must also be acknowledged that, unlike surveys, administrative data cannot capture factors such as perceptions, beliefs, attitudes, knowledge, and reasoning. Similarly, while economists often favor the revealed preference approach, many crucial determinants of social, economic, and political outcomes – such as perceptions, beliefs, attitudes, and reasoning – are not always easily inferred from observed behavior. This limitation makes surveys a valuable complement to both administrative and other observational data. This is especially relevant in the context of discrimination, where identifying its sources solely through aggregate statistics and observational data can be challenging (Domínguez, Grau, and Vergara 2022).

(Stantcheva 2023) also specifically highlighted some advantages of online surveys in terms of selection, as compared to in-person, phone, or mail surveys. First, online surveys give people the flexibility to complete the survey at their convenience, which reduces selection based on who is free to answer during regular work hours or who opens the door or picks up the phone. This feature may allow individuals who need to juggle different responsibilities (e.g., carers) to take part in a study. Second, the convenience of mobile technologies may entice some people who would otherwise not want to fill out questionnaires or answer questions on the phone to take surveys. Third, online surveys can reach people who would be hard to interview in person (e.g., younger respondents, those who often move residences, respondents in remote or rural areas, etc.). Fourth, platforms administering this kind of survey offer a variety of rewards for taking surveys, which can appeal to a broader group of people.

**A.5 List experiments assumptions**

The validity of a list experiment relies on three assumptions: 1) treatment randomization, 2) no design effect, and 3) 'no liar'. The first assumption means that the sample is split at random. The second assumption means that respondents do not give different answers to non-key statements depending on whether they are in the long list group. The third assumption means that respondents answer the key statement truthfully.

A common practice to check the first assumption – treatment randomization – is to test for differences between the short list and long list groups' responses to important variables in the survey. More precisely, since this study is based on double list experiments, one has to check whether participants treated in lists A are systematically different than participants treated in lists B. Table A1 checks the differences between the two groups in terms of their demographic covariates. Largely, there are no significant differences between the two groups: except for slight differences (marginally statistically significant and negligible in magnitude) for the age group 50-64 and for individuals making more than approximately $975 USD. This evidence is thus reassuring that the randomization of treatment was effective.

Moreover, following (Gerber and Green 2012) and (Detkova, Tkachenko, and Yakovlev 2021), the main analysis does not only rely on means comparisons but also employs regression analyses controlling for observable characteristics (as discussed in Section 3.1), including subjects' age, sex at birth, sexual orientation, gender identity, indicators for ethnicity (i.e., African descent) and indigenous status, region and commune (for those in the Metropolitan region of Santiago) fixed effects, education level, income, employment status, an indicator for managerial experience, current religious affiliation, and political affiliation. Additional controls include whether at least one child less than 18 years of age lives in the subject's household, number of people living in the subject's household, marital status, LGBTQ+ comfort (as defined in Appendix B), an indicator for whether the participant knows a gay person (as defined in Appendix B), and day-of-week and week-of-sample indicators.

The second assumption – no design effect – requires respondents not to change their answers to non-key statements depending on whether the key statement appeared in the list (i.e., whether they saw the long list). To clarify, suppose that a respondent in the short list group answered two non-key statements affirmatively. If they had been assigned to the long list group, their answer must have been either '2' or '3' (that is, they either answered two non-key statements affirmatively or they answered two non-key statements plus the key statement affirmatively). It is worth noting that it is not assumed that subjects gave truthful answers to these non-key statements, it is only assumed that the answers were consistent in short and long list groups. (Blair and Imai 2012) proposed a statistical test for the no design effect assumption. This can be implemented using the Stata command *kict deff* (Tsai 2019). The first step is to estimate the probabilities of all possible types of item count responses. If some of these estimated probabilities were a nonsensical value (e.g., a negative value), it would raise doubts about the validity of the no-design-effect assumption. One can then test whether such negative estimates have arisen by chance.

Following (Blair and Imai 2012) and (Tasi 2019), we perform the no design effect tests twice, with and without using the method of generalized moment selection (GMS). Using GMS, 6 (out of 60) estimated probabilities of the item count responses present negative values, and 2 (out of 12) Bonferroni-adjusted p-values are slightly above conventional significance thresholds. These marginally significant p-values correspond to List B (Supervisor) and List A (Cashier). Conducting the same tests without using the GMS method, we find that the same number of estimated probabilities of the item count responses present negative values, but all Bonferroni-adjusted p-values are above conventional significance thresholds. That is, all tests fail to reject the null hypothesis of no design effects. That is, one cannot reject the null that such negative estimates have arisen by chance. Overall, the results indicate that there is no strong evidence of bias in the list experiment design across conditions. Therefore, it is possible to conclude that the available evidence supports the "no design effect" assumption.

It is not statistically feasible to check the 'no liar' assumption, not only because respondents' answers to the key statement are by design unobserved, but also because their truthful answers are unknown (otherwise there would be no point in using the list experiment technique). This study tried to limit any concerns about this assumption by running these experiments in an online anonymized platform and by making sure when designing the lists that agreeing to all or none of the statements is highly unlikely. Indeed, Figures A1-A3 present the distribution of responses for each list and key statements: the modal response in most lists is 2. Moreover, as noted in the previous section, the percentage of times where the responses are 0 or 4 (5 for long lists) is negligible, meaning that the privacy of responses was protected.

## A.6. Data quality checks

### A.6.1 Procrastinators and speeders.

As also done by (Gutiérrez and Rubli 2024), it is possible to check the robustness of the main list experiment findings by excluding participants who completed the study very quickly or very slowly since they may not be paying as much attention to the study instructions. The median respondent took 848 seconds (14.1 minutes) to complete the list experiments. The results presented in Figure A4 show that the main findings are robust to excluding 401 participants who took less than 491 seconds (or 8.1 minutes, top 5%) or more than 3354.5 seconds (or 55.9 minutes, bottom 5%).

**A.6.2 Attention checks.**

The list experiments included an attention check, and the survey included two additional attention check questions. Following the recommendation in (Haaland, Roth, and Wohlfart 2023), respondents were explained at the beginning of the study the rationale for including such attention checks. This explanation can mitigate concerns about participants' negative emotional reactions to the use of attention checks. More specifically, respondents were informed that sometimes there are participants who do not carefully read the questions and just quickly click through surveys, thus resulting in random answers which compromise the research findings. For this reason, they were told that the study included several attention checks and that failing to complete two or more of these questions correctly may cause them not to be eligible for compensation.

776 participants (19.4% of our sample) failed one out of the three attention checks. In particular, 761 of these 776 participants failed the first attention check, arguably the most challenging one. 0.88% percent (n= 35) of the participants failed two or more attention checks. For robustness, Figure A5 presents results excluding participants who failed at least one attention check. Our results are robust to the exclusion of these participants.

It is also possible to check if some respondents provided the same answer for all list experiments (which might be an indication of participants not paying attention). Across all seven lists, 12 participants (0.3% of our sample) provided the same number. Looking at the first six lists (thus excluding the list that serves as an attention check), 58 participants (1.45%) provided the same number for all four lists. As shown in Figure A6, our main findings are robust to the exclusion of these 58 participants.

It is also worth mentioning that, as suggested in (Stantcheva 2023) to verify that respondents were humans, all participants were required to complete a CAPTCHA test.

In addition, the Qualtrics software collected basic geographic information to verify that all respondents were indeed taking the survey from Chile. Figure A7 presents our main findings excluding 98 participants who completed our study from outside of Chile according to Qualtrics approximation of the respondent's latitude and longitude and 62 participants for whom Qualtrics did not provide a location approximation. Results are robust to the exclusion of these responses.

**A.6.3 Attrition rate**

As stated in the consent form, subjects were always free to leave the study whenever they wished. The attrition rate was 23.98%, that is, 1,251 participants started the study but did not complete it. This attrition occurs overwhelmingly *before* the start of the list experiment. Specifically, 775 participants (61.95% of all dropouts) exited our study before the start of the list experiment (i.e., during consent, overview, instructions or the example list experiment questions). Out of the remaining participants, 144 (11.5%) dropped out *during* the list experiment. The remaining respondents who went on to complete the survey, did not seem to display a pattern of abandoning the study more frequently after specific survey questions. Figure A8 plots the number of non-missing values for variables at specific points through the study. This figure serves to illustrate the patterns of attrition mentioned above, namely that dropouts occurred mostly *before* the start of the list experiment. It is also worth noting that, without considering individuals who did not start the list experiments, the attrition rate was only 10.63%, thus much lower than surveys conducted by national statistical offices.

Our main sample only uses data from participants who completed the entire study. The main sample of 4,000 respondents does not include any of the 1,251 participants described above, who dropped out before finishing our study.

**A.6.4 Feedback questions.**

At the end of the survey, participants were asked to report the extent to which they agreed with the statement "The instructions were clear". The menu of answers consisted of a 5-point Likert scale ranging from "Totally agree" to "Totally disagree". As Table A2 shows, over 97% of participants indicated total or somewhat agreement with the statement, suggesting that most respondents found the instructions clear. However, for robustness, Figure A9 shows the main list results excluding 63 participants who responded "Neither agree nor disagree", 25 participants who responded "Somewhat disagree" and 16 participants who responded "Totally disagree" with the statement "The instructions were clear". Results remain robust to the exclusion of these 104 respondents.

This 5-point Likert scale question was followed by two open-ended questions to provide respondents with a venue to leave feedback about the study to the researchers. The first open-ended question asked respondents to report anything that may have not been clear or that could have been confusing during the study. 2,385 participants provided responses to the prompt "Is there anything that is not clear or is confusing in the study?". Responses were processed by first cleaning the text (converting to lowercase, removing punctuation, and filtering out stopwords) and then categorized into thematic groups. This analysis reveals that over half of the respondents (56.23%) indicated "No Issues," suggesting that the study was largely perceived as clear. Additionally, 205 participants (8.60%) offered explicit "Positive Feedback" about the study's clarity. A small fraction (1.38%) raised "Clarity Issues," while 0.80% provided "Suggestions" for improvement, and 13.50% of responses fell into the "Mixed" category, reflecting overlapping or ambiguous feedback. Finally, 19.50% of the responses were "Uncategorized," as they did not neatly align with any of the predefined themes. Among those who raised issues, their chief complaint was about the purpose of the study: an additional manual inspection of a randomly selected subset of 100 respondents indicates that several participants were confused about the objective of the study and the reason why the list experiment questions were asked. These comments are in line with our objective, as we did not want to prime respondents by disclosing the aim of the survey.

The second and last open-ended question asked participants if there was anything else they wanted to share with the investigators. A similar analysis found that, out of 2,307 non-missing responses to the prompt "Is there anything else that you would like to share with the investigators?", the majority (53.75%) simply provided "No Comments," indicating they had nothing further to share. A substantial share of responses (37.62%) were "Uncategorized," meaning they did not neatly fit into the predefined thematic patterns. In contrast, smaller proportions conveyed distinct sentiments: 1.73% were flagged as "Negative," 3.12% as "Positive," 2.60% offered "Suggestions" for improvement, and 1.17% were classified as "Mixed." Most respondents offered minimal additional feedback in this question and inspections of a randomly selected subset of 100 responses uncovered the diversity of comments, ranging from questions about specific portions of the survey to personal anecdotes related to the content of our study.

Overall, analysis of these final survey questions does not reveal any worrying signs about the survey design and data quality.

### A.6.5 Standard errors and weights

Standard errors were computed following (Glynn 2013): because estimation is accomplished by taking the difference in mean responses between two independent sets of respondents, the variance of the estimator can be calculated with the standard large-sample formula for a difference in means, and confidence intervals can be computed in the usual fashion.

Figure A10 further shows that the main findings are robust using unweighted data, although the estimated share of people comfortable with a gay cashier from the list experiments is higher in the unweighted data.

### A.6.7 Datavoz Pilot and Soft Launch

After we piloted the experiment on Prolific with a sample of Chilean participants (discussed in detail in Appendix C), our local partner, Datavoz, conducted a pilot and soft launch of our study with the same target population as our main sample, before launching our main data collection effort. The Datavoz pilot consisted of 235 responses collected in September 2024 and the soft launch consisted of 62 responses collected in October 2024. The purpose of these exercises was to test our instrument in the field, gather initial feedback from participants, and adjust our instrument as needed.

As outlined in our pre-analysis plan, Figure A11 shows our main list results with the inclusion of data from both the pilot and soft launch, with a total of 4,297 observations. Unweighted results including these additional samples largely mimic those presented in Figure 1, although with some notable differences. In particular, while Figure 1 shows slightly higher estimates of social desirability bias for supervisors and customers, Figure AX presents slightly higher estimates for co-workers. The main difference across the figures stems from higher levels of support estimated from the double list experiment from customers.

### A.7.7 Additional considerations.

Before pre-registering the experiment and starting the data collection, the pre-analysis plan and survey were reviewed by four experts on list experiments. The questionnaire and experiment were then modified to incorporate their recommendations.

Finally, we chose to rely on list experiments instead of the randomized response technique (where respondents use a private randomization device - e.g., flip a coin - to determine whether they answer either a sensitive or innocuous question) for three reasons. First, the randomized response technique is more difficult to implement online. Second, research suggests that subjects trust the randomized response technique less than the list experiment (Coutts and Jann 2011). Third, research also documents that participants may not respond to the randomization device relied upon by the this technique as instructed (John et al. 2018). For these reasons, and for the advantages of our methods discussed in Section A.4.7, we decided to rely on double list experiments.

**Figure A1. Responses to the Lists Supervisor with and Without the Supervisor Key Statement**

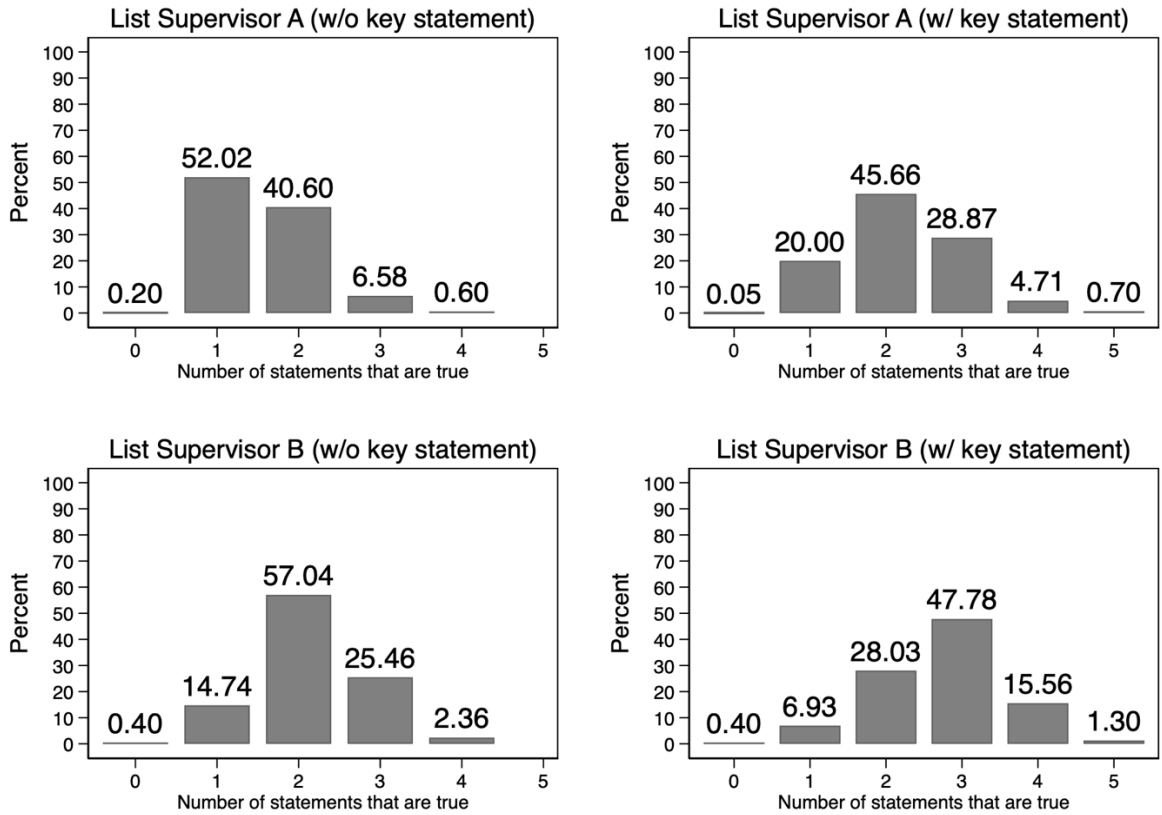

*Notes:* Unweighted statistics. Key statement: "I would feel comfortable supervising a gay employee." Number of observations: 2,005 (List A without key); 1,995 (List A with key); 1,995 (List B without key); 2,005 (List B with key).

**Figure A2. Responses to the Lists Co-worker with and without the Co-worker Key Statement**

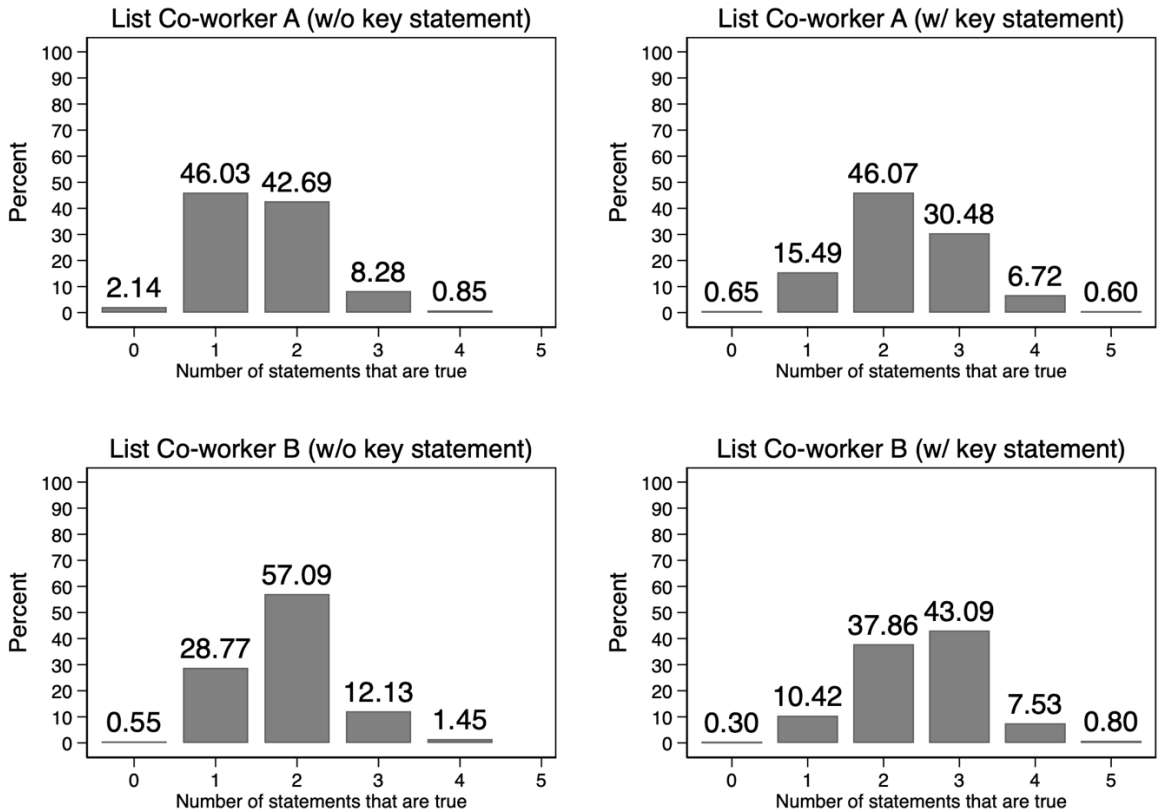

*Notes:* Unweighted statistics. Key statement: "I would feel comfortable working closely with a gay co-worker." Number of observations: 2,005 (List A without key); 1,995 (List A with key); 1,995 (List B without key); 2,005 (List B with key).

**Figure A3. Responses to the Lists Cashier with and without the Customer Key Statement**

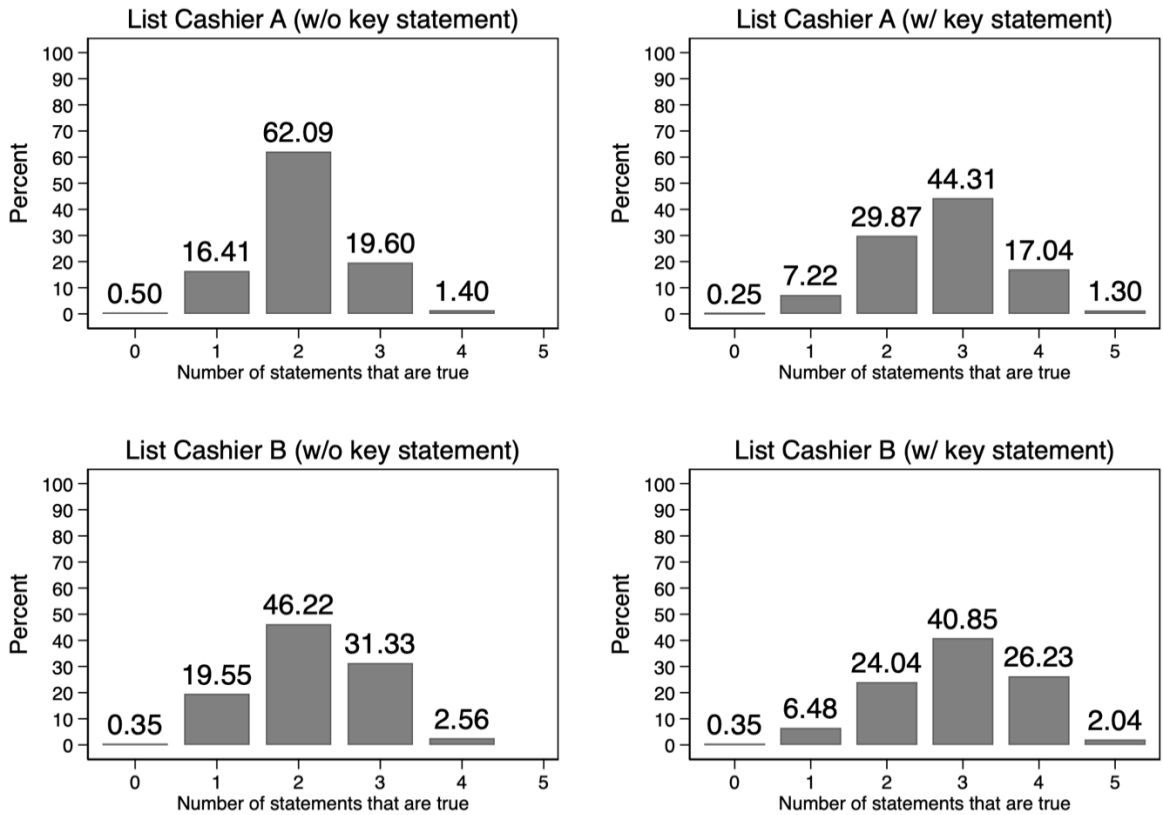

*Notes:* Unweighted statistics. Key statement: "I would feel comfortable having a cashier at the supermarket who is gay." Number of observations: 2,005 (List A without key); 1,995 (List A with key); 1,995 (List B without key); 2,005 (List B with key).

**Figure A4. List Experiment on Attitudes from Gay Supervisors, Co-workers, and Customers – Excluding Speeders and Procrastinators**

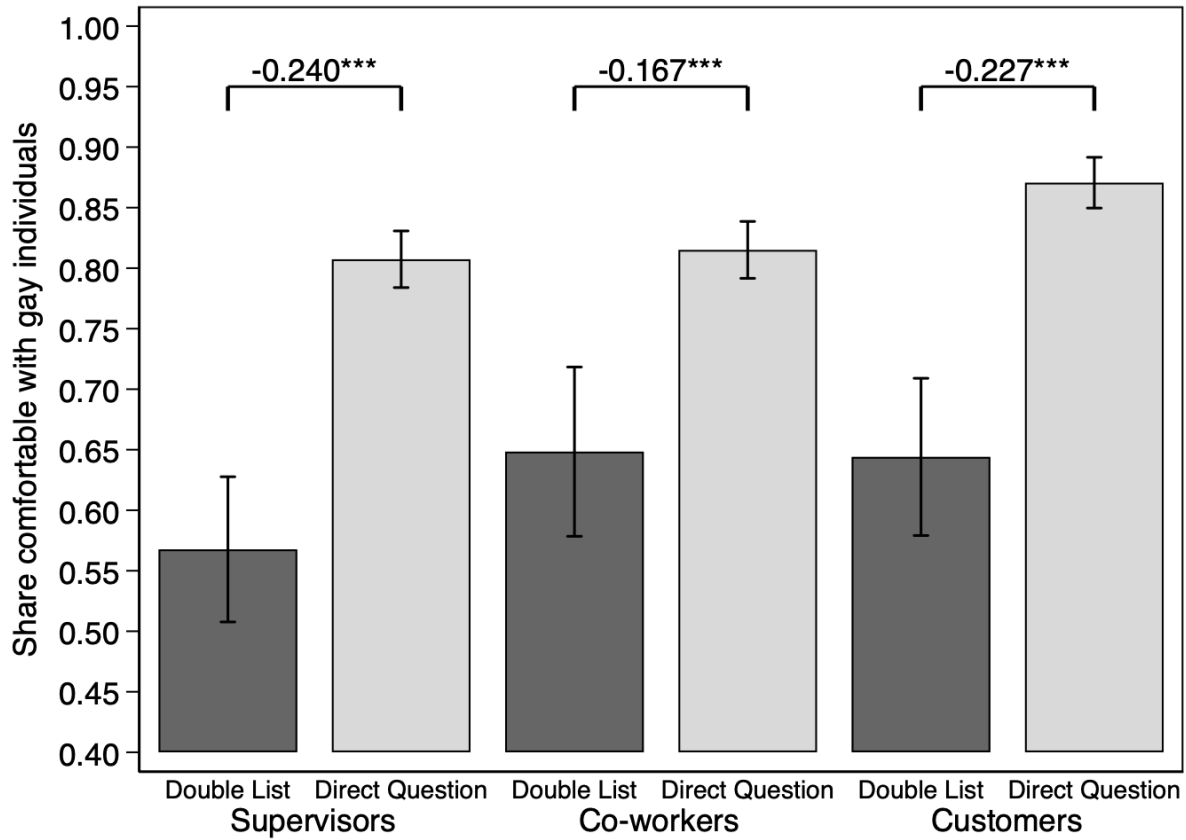

*Notes:* Weighted statistics. 95% confidence intervals are reported with the vertical range plots. The numbers above the horizontal bars are the differences between the two groups at the base of each horizontal bar. Supervisor key statement: "I would feel comfortable supervising a gay employee." Co-workers key statement: "I would feel comfortable working closely with a gay colleague." Customers key statement: "I would feel comfortable having a gay cashier at the supermarket." Number of observations: 3,599. Sample does not include 401 participants who took less than 491 seconds (or 8.1 minutes - top 5%) or more than 3354.5 seconds (or 55.9 minutes - bottom 5%). *$p < 0.10$; **$p < 0.05$; ***$p < 0.01$.

**Figure A5. List Experiment on Attitudes from Gay Supervisors, Co-workers, and Customers - Excluding Participants Who Failed One or More Attention Checks**

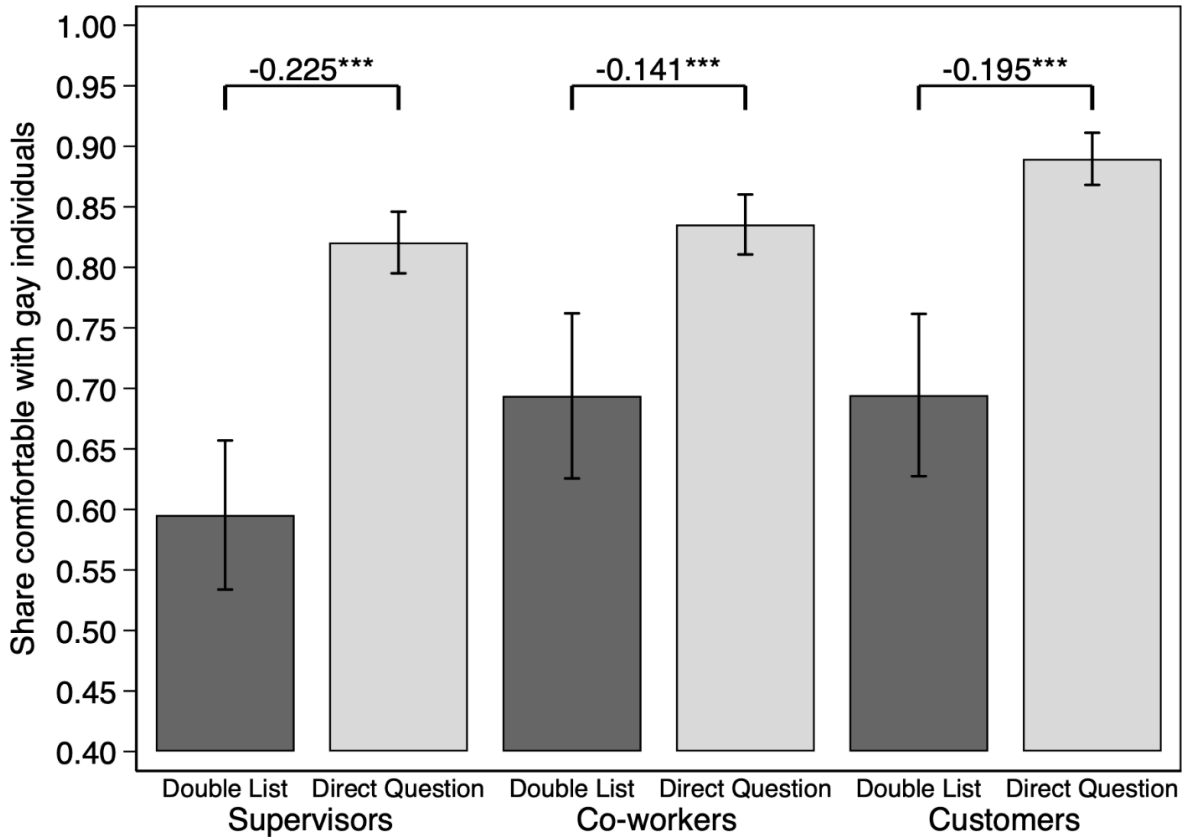

*Notes:* Weighted statistics. 95% confidence intervals are reported with the vertical range plots. The numbers above the horizontal bars are the differences between the two groups at the base of each horizontal bar. Supervisor key statement: "I would feel comfortable supervising a gay employee." Co-workers key statement: "I would feel comfortable working closely with a gay colleague." Customers key statement: "I would feel comfortable having a gay cashier at the supermarket." Number of observations: 3,224. Sample does not include 776 participants who failed at least one attention check. *p < 0.10; **p < 0.05; ***p < 0.01.

**Figure A6. List Experiment on Attitudes from Gay Supervisors, Co-workers, and Customers - Excluding Participants Who Provided a Constant Answer in All Six Lists**

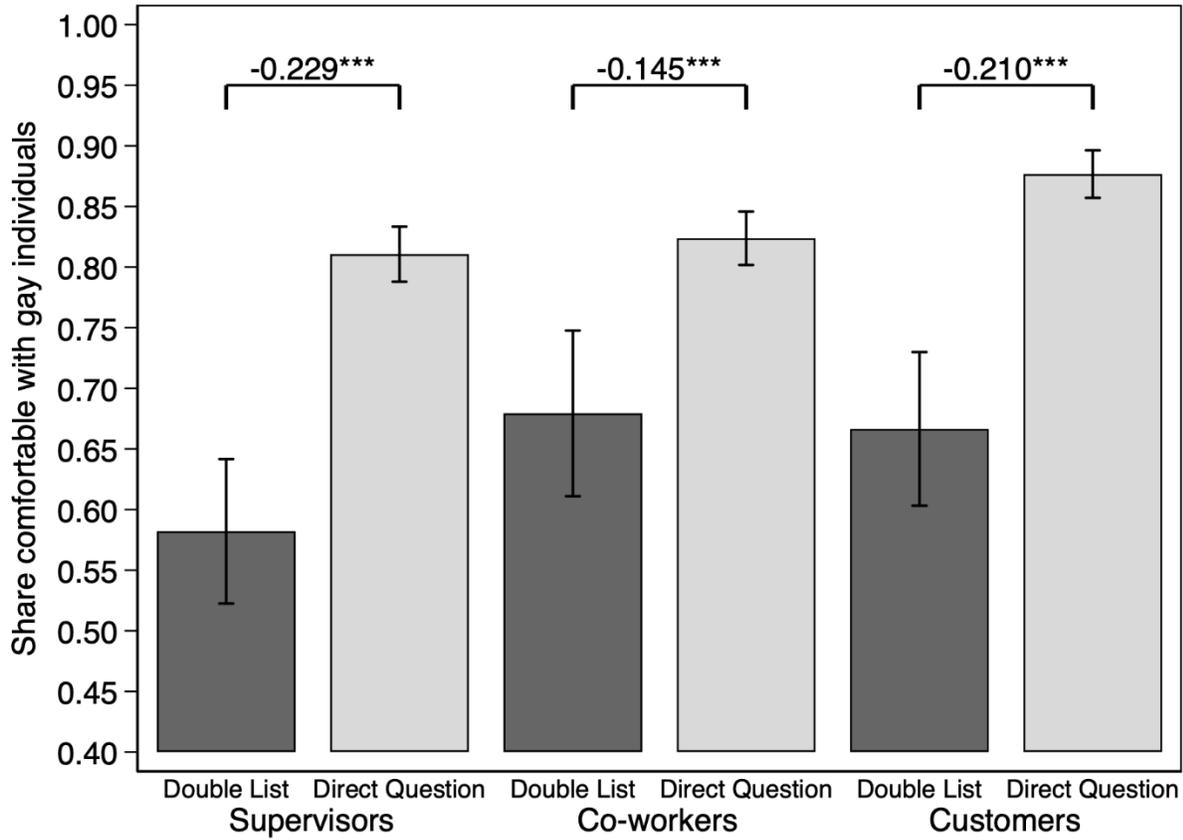

*Notes:* Weighted statistics. 95% confidence intervals are reported with the vertical range plots. The numbers above the horizontal bars are the differences between the two groups at the base of each horizontal bar. Supervisor key statement: "I would feel comfortable supervising a gay employee." Co-workers key statement: "I would feel comfortable working closely with a gay colleague." Customers key statement: "I would feel comfortable having a gay cashier at the supermarket." Number of observations: 3,942. Sample does not include 58 participants who provided the same response in all six lists. *p < 0.10; **p < 0.05; ***p < 0.01.

**Figure A7. List Experiment on Attitudes from Gay Supervisors, Co-workers, and Customers - Excluding Participants Who Responded Outside of Chile or Did Not Have Location Data**

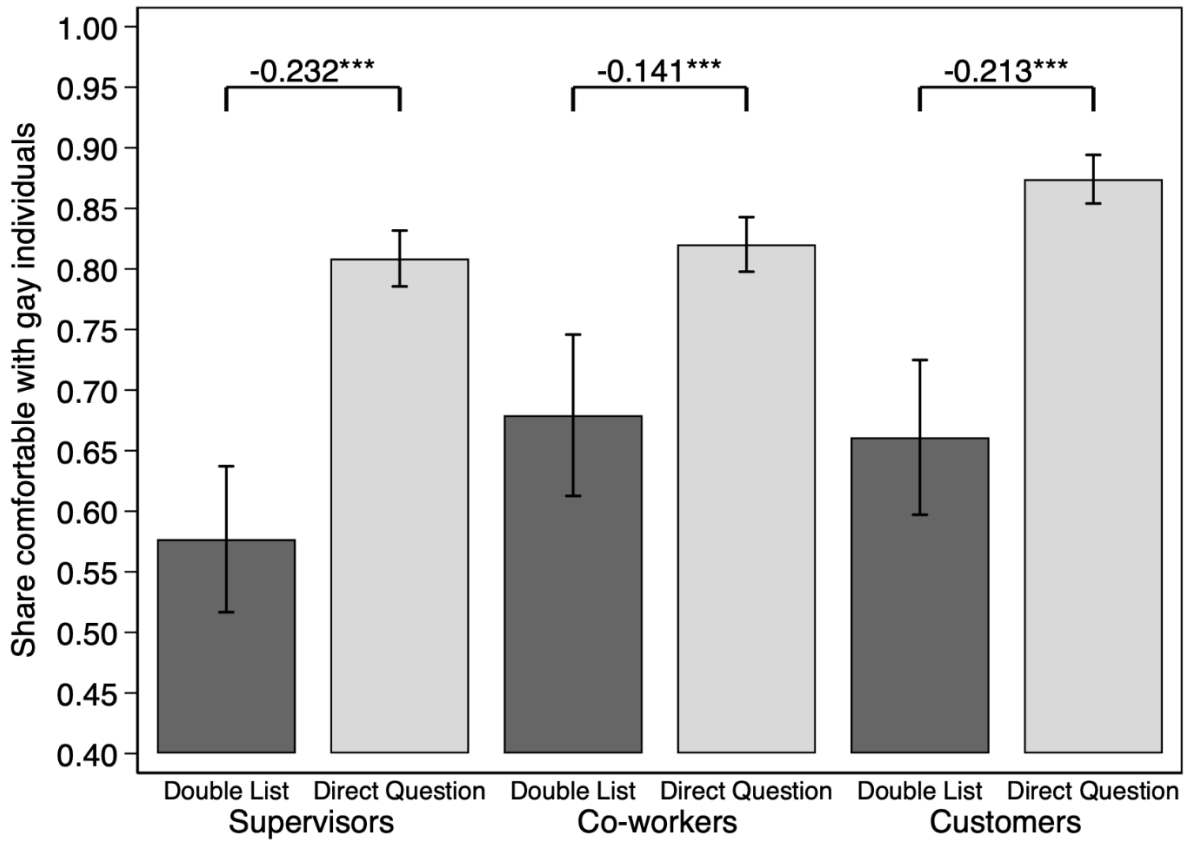

*Notes:* Weighted statistics. 95% confidence intervals are reported with the vertical range plots. The numbers above the horizontal bars are the differences between the two groups at the base of each horizontal bar. Supervisor key statement: "I would feel comfortable supervising a gay employee." Co-workers key statement: "I would feel comfortable working closely with a gay colleague." Customers key statement: "I would feel comfortable having a gay cashier at the supermarket." Number of observations: 3,840. Sample does not include 98 participants who responded to our survey from outside of Chile and 62 participants for whom Qualtrics did not provide location data. *p < 0.10; **p < 0.05; ***p < 0.01.

**Figure A8. Attrition Numbers by Survey Question**

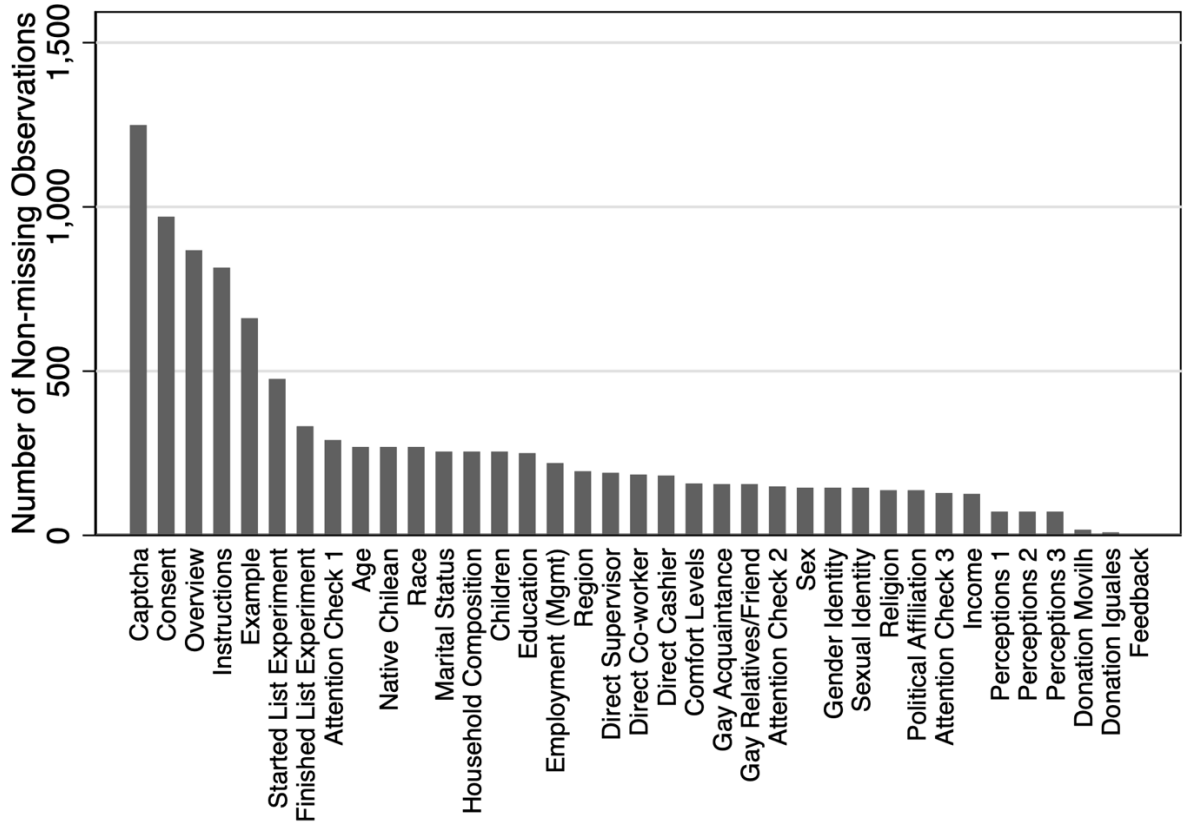

*Notes:* Unweighted statistics. Number of starting observations: 1,251. The variables listed in the horizontal axis appear sequentially, following the order in which they were presented in our study.

**Figure A9. List Experiment on Attitudes from Gay Supervisors, Co-workers, and Customers - Excluding Participants Who Did Not Agree with "The instructions were clear" Statement**

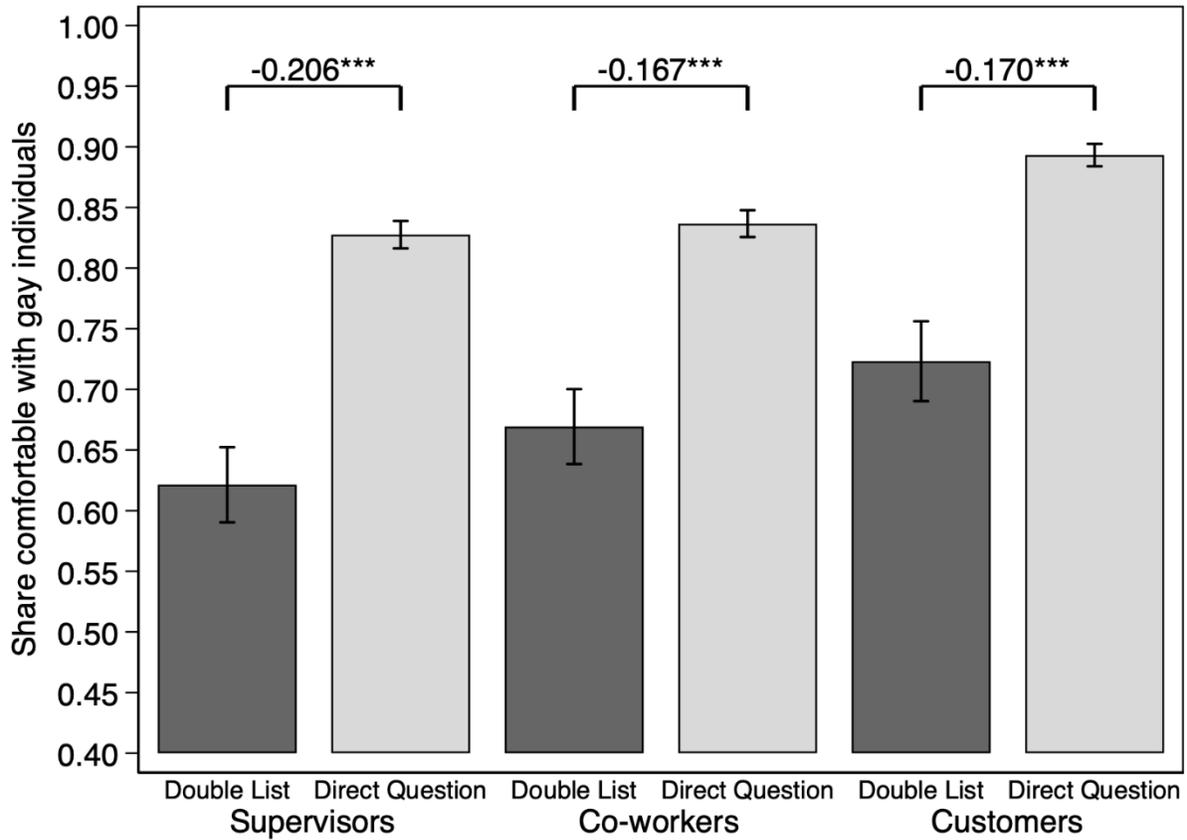

*Notes:* Weighted statistics. 95% confidence intervals are reported with the vertical range plots. The numbers above the horizontal bars are the differences between the two groups at the base of each horizontal bar. Supervisor key statement: "I would feel comfortable supervising a gay employee." Co-workers key statement: "I would feel comfortable working closely with a gay colleague." Customers key statement: "I would feel comfortable having a gay cashier at the supermarket." Sample does not include 63 participants who responded "Neither agree nor disagree", 25 participants who responded "Somewhat disagree" and 16 participants who responded "Totally disagree" with the statement "The instructions were clear". Number of observations: 3,896. *p < 0.10; **p < 0.05; ***p < 0.01.

**Figure A10. List Experiment on Attitudes from Gay Supervisors, Co-workers, and Customers - Unweighted Data**

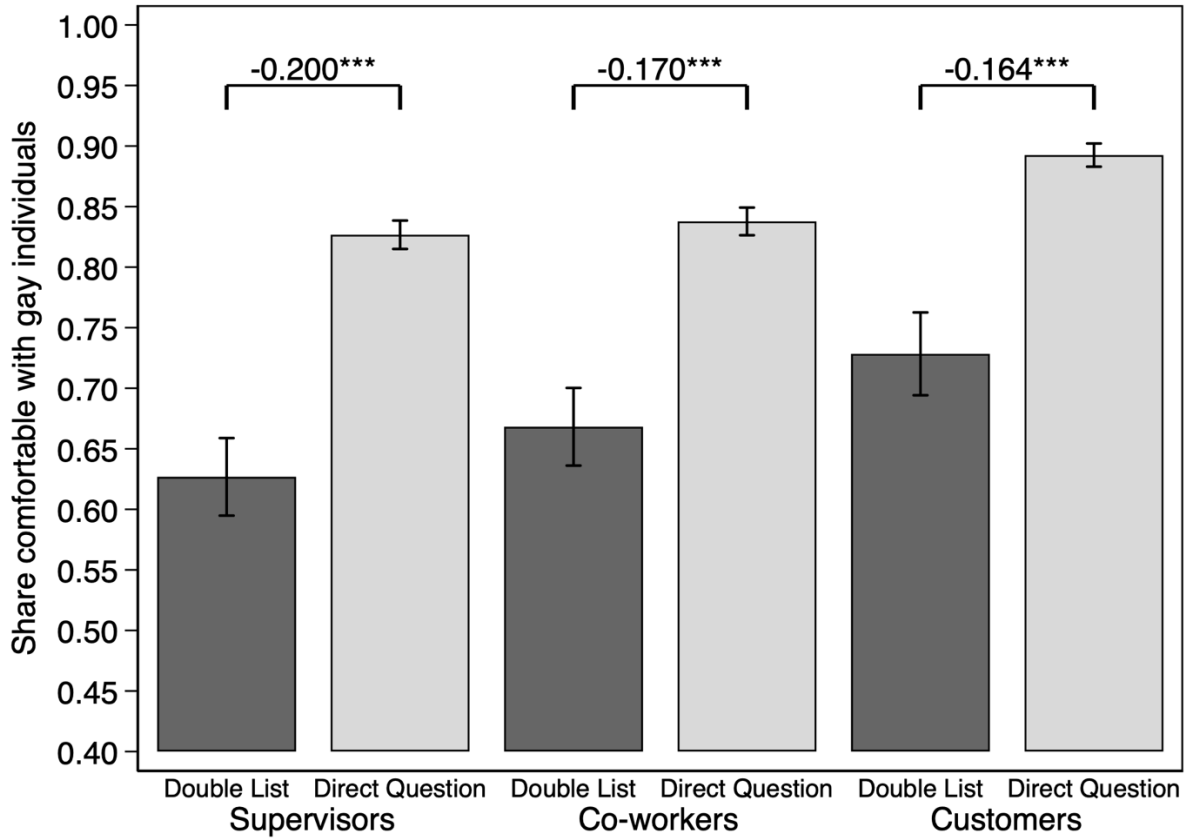

*Notes:* Unweighted statistics. 95% confidence intervals are reported with the vertical range plots. The numbers above the horizontal bars are the differences between the two groups at the base of each horizontal bar. Supervisor key statement: "I would feel comfortable supervising a gay employee." Co-workers key statement: "I would feel comfortable working closely with a gay colleague." Customers key statement: "I would feel comfortable having a gay cashier at the supermarket." Number of observations: 4,000. *p < 0.10; **p < 0.05; ***p < 0.01.

**Figure A11. List Experiment on Attitudes from Gay Supervisors, Co-workers, and Customers - Including Data from Datavoz Pilot and Soft Launch**

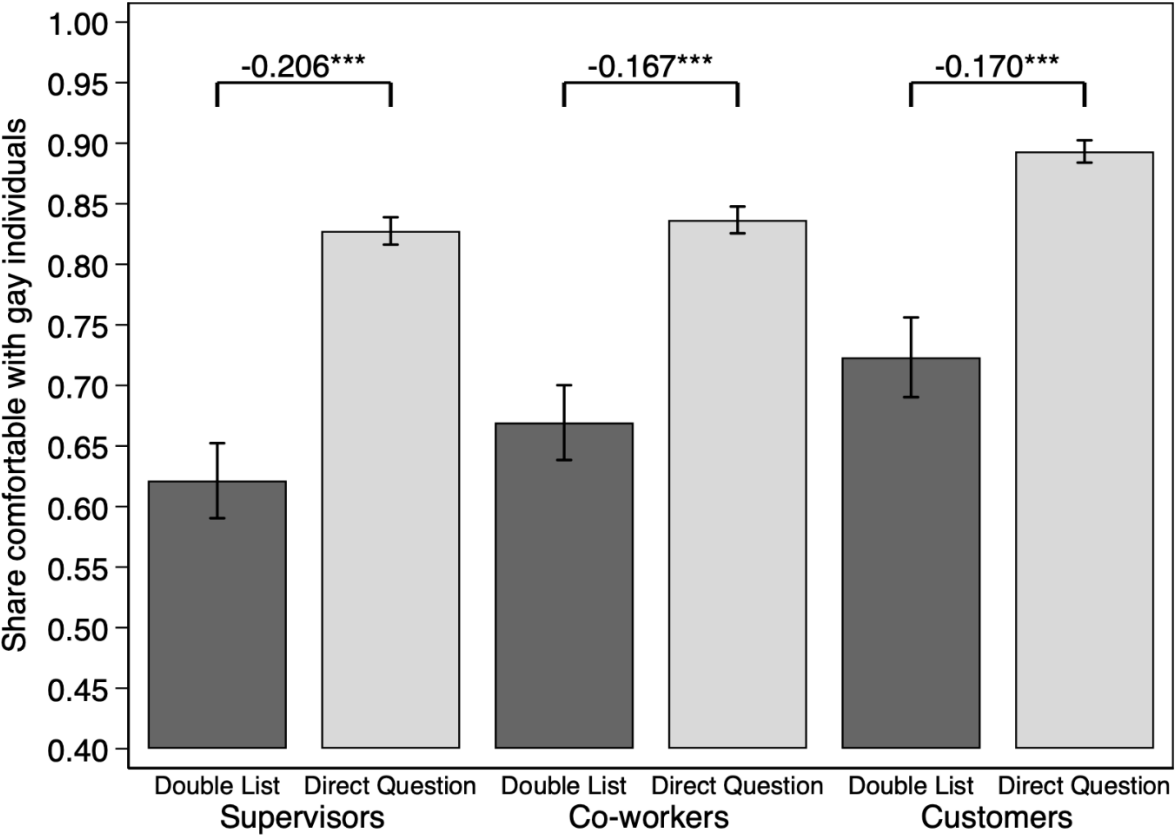

*Notes:* Unweighted statistics. 95% confidence intervals are reported with the vertical range plots. The numbers above the horizontal bars are the differences between the two groups at the base of each horizontal bar. Supervisor key statement: "I would feel comfortable supervising a gay employee." Co-workers key statement: "I would feel comfortable working closely with a gay colleague." Customers key statement: "I would feel comfortable having a gay cashier at the supermarket." Sample includes data from a first pilot (N=235) conducted in September 2024 and a soft launch (N=62) conduced in October 2024, and our main wave (N=4,000) conducted in November – December 2024. Total number of observations: 4,297. *p < 0.10; **p < 0.05; ***p < 0.01.

**Table A1. Balance Table**

| Variable | Treatment A Mean | Treatment B Mean | Difference | p-value |
|---|---|---|---|---|
| Age | 47.61 | 47.77 | 0.16 | 0.678 |
| *Between 18 and 34* | 0.131 | 0.125 | -0.006 | 0.594 |
| *Between 35 and 49* | 0.477 | 0.457 | -0.019 | 0.221 |
| *Between 50 and 64* | 0.280 | 0.313 | 0.033 | 0.022** |
| *65 or older* | 0.113 | 0.105 | -0.008 | 0.414 |
| Sex assigned at birth: Female | 0.477 | 0.463 | -0.014 | 0.380 |
| African Descent | 0.032 | 0.027 | -0.004 | 0.438 |
| Indigenous | 0.067 | 0.066 | -0.001 | 0.916 |
| Gender Identity and Sexual Orientation | | | | |
| *Cisgender* | 0.992 | 0.994 | 0.002 | 0.552 |
| *Heterosexual* | 0.926 | 0.926 | -0.000 | 0.995 |
| Income: More than $975 | 0.614 | 0.642 | 0.028 | 0.064* |
| Education | | | | |
| *High School* | 0.181 | 0.168 | -0.014 | 0.248 |
| *Bachelor's Degree* | 0.470 | 0.489 | 0.019 | 0.227 |
| *Post-Graduate Degree* | 0.219 | 0.223 | 0.004 | 0.767 |
| Employed | 0.847 | 0.852 | 0.005 | 0.674 |
| Management Experience | 0.771 | 0.767 | -0.004 | 0.773 |
| Region: Non-Metro | 0.366 | 0.355 | -0.011 | 0.477 |
| Political Affiliation | | | | |
| *Leans Right* | 0.460 | 0.454 | -0.006 | 0.704 |
| *Leans Left* | 0.540 | 0.546 | 0.006 | 0.704 |
| Religious Affiliation | | | | |
| *Catholic* | 0.490 | 0.483 | -0.007 | 0.644 |
| *No Religion* | 0.365 | 0.376 | 0.011 | 0.483 |
| LGBTQ+ Comfortable | 0.975 | 0.982 | 0.006 | 0.172 |
| Has Gay Relative(s) or Friend(s) | 0.806 | 0.798 | -0.008 | 0.552 |

*Notes:* Weighted statistics. Treatment A refers to participants randomized into seeing the key sensitive statements in List A as opposed to List B. Treatment B refers to participants randomized into seeing the key sensitive statements in List B as opposed to List A.

**Table A2: Instructions feedback**

| Indicate the extent to which you agree or disagree with the following statement: "The instructions were clear" | Number of observations | Share of respondents |
| --- | --- | --- |
| Totally agree | 3,582 | 89.55 |
| Somewhat agree | 314 | 7.85 |
| Neither agree nor disagree | 63 | 1.57 |
| Somewhat disagree | 25 | 0.62 |
| Totally disagree | 16 | 0.40 |

Note: the original Spanish question is "Indique en qué medida está de acuerdo o en desacuerdo con la siguiente afirmación: 'Las instrucciones fueron claras'."

**Appendix B. Variables description.**

Unless otherwise noted, all respondents were required to provide an answer to every question before being allowed to proceed further. As noted below, some questions allowed respondents to select options such as "Do not know" or "Prefer not to answer". Summary statistics for all variables are provided in Table B1.

*†* Denotes questions with similar or exact language sourced from the 2024 Census of Chile questionnaire (Cuestionario de Viviendas Particulares, Censo de Poblacion y Vivienda 2024).

*‡* Denotes questions with similar or exact language sourced from the 2018 World Values Survey (WVS7 Questionnaire Chile 2018).

*§* Denotes questions with similar or exact language sourced from the 2020 Latinobarometro Questionnaire.

- *Age†* reports the respondent's age in year.
- *Indigenous†* indicates whether the respondent belongs or identifies as belonging to an indigenous or native people group.
- *Race†* reports whether the respondent, according to their ancestors, traditions and culture, is or considers themselves of African descent/ancestry.
- *Marital Status†* reports the respondent's current marital status.
- *Household Composition‡* reports the number of people who live as members of the respondent's household, including children.
- *Children‡* indicates whether the respondent has children.
- *Education†* reports the highest educational level achieved by the respondent.
- *Employment†* reports the employment status of the respondent during the previous week.
- *Employment Role†* reports the role (e.g., employer, employee, domestic worker, etc.) of the respondent (if employed) during the previous week.
- *Employment Management* indicates whether the respondent has had any work experience as a supervisor of one or more workers
- *Employment Contract* indicates whether the respondent has a written contract in their main job or economic activity.
- *Employment Report* indicates whether the respondent normally reports income from their job or main economic activity to any government entity.
- *Occupation§* reports the occupational group of the respondent in general, independently of whether the respondent was employed the previous week.
- *Number (of) Colleagues* reports the number of people (including colleagues, bosses, and employees) that respondents typically interact with in a week at their current job (if employed).
- *Region* reports the region of Chile the respondent currently lives in.
- *Commune* reports the commune the respondent currently lives in (only if the participant indicated Santiago Metropolitan Region in the previous question).
- *Birth Country* reports the country of birth of the respondent.
- *Sex†* indicates the sex assigned at birth (male/female) of the participant.

- *Gender Identity[†]* reports the gender the respondent identifies with (includes Don't know/Prefer not to answer options).
- *Sexual Identity* reports the sexual orientation the respondent identifies with (includes Don't know/Prefer not to answer options).
- *Religion[†]* reports the respondent's religion (includes Prefer not to answer option).
- *Political Affiliation[‡]* reports the respondent's political affiliation on a 1-10 scale, with 1 denoting "left" and 10 denoting "right" (response is not required).
- *LGBTQ+ Comfort* takes value 1 if the respondent reports that they would maintain the same relationship or even become closer with an acquaintance if they were to reveal to the respondent that they are gay.
- *LGBTQ+ relatives or friends* takes value 1 if the respondent reports knowing someone who is gay among their immediate family, relatives, neighbors, co-workers or friends.

## Table B1: Descriptive Statistics

| Variable | N | Mean | SD | Min | Max |
|---|---|---|---|---|---|
| Age | 4,000 | 45.93 | 14.06 | 18 | 87 |
| *Between 18 and 34* | 4,000 | 0.320 | 0.467 | 0 | 1 |
| *Between 35 and 49* | 4,000 | 0.274 | 0.446 | 0 | 1 |
| *Between 50 and 64* | 4,000 | 0.295 | 0.456 | 0 | 1 |
| *65 or older* | 4,000 | 0.111 | 0.314 | 0 | 1 |
| Sex assigned at birth: Female | 4,000 | 0.511 | 0.500 | 0 | 1 |
| African Descent | 4,000 | 0.049 | 0.215 | 0 | 1 |
| Indigenous | 4,000 | 0.101 | 0.301 | 0 | 1 |
| Gender Identity and Sexual Orientation | | | | | |
| *Cisgender* | 3,990 | 0.991 | 0.094 | 0 | 1 |
| *Heterosexual* | 3,952 | 0.920 | 0.271 | 0 | 1 |
| Income: More than $975 | 3,838 | 0.602 | 0.490 | 0 | 1 |
| Education | | | | | |
| *High School* | 4,000 | 0.406 | 0.491 | 0 | 1 |
| *Bachelor's Degree* | 4,000 | 0.194 | 0.396 | 0 | 1 |
| *Post-Graduate Degree* | 4,000 | 0.076 | 0.265 | 0 | 1 |
| Employed | 4,000 | 0.806 | 0.395 | 0 | 1 |
| Management Experience | 4,000 | 0.667 | 0.471 | 0 | 1 |
| Region: Non-Metro | 4,000 | 0.578 | 0.494 | 0 | 1 |
| Political Affiliation | | | | | |
| *Leans Right* | 3,763 | 0.445 | 0.497 | 0 | 1 |
| *Leans Left* | 3,763 | 0.555 | 0.497 | 0 | 1 |
| Religious Affiliation | | | | | |
| *Catholic* | 3,899 | 0.499 | 0.500 | 0 | 1 |
| *No Religion* | 3,899 | 0.320 | 0.467 | 0 | 1 |
| LGBTQ+ Comfortable | 3,742 | 0.976 | 0.153 | 0 | 1 |
| Has Gay Relative(s) or Friend(s) | 4,000 | 0.777 | 0.417 | 0 | 1 |

*Notes*: Weighted statistics. Missing observations reflect cases in which respondents did not respond to a question or selected options such as "I prefer not to respond".

**Appendix C. Prolific pilot.**

Individuals from Chile were recruited on Prolific before Datavoz started any data collection in order to pilot the list experiments and survey. The Qualtrics questionnaire used with the Prolific participants is the same as the one used with the Datavoz participant and reported in Appendix E.3, with a few important exceptions. First, as analyzed in Section C.1, only half of the participants saw the list experiments, while the other half only took part in the survey. Second, as analyzed in Section C.2, Prolific participants were asked additional questions in order to make sure that the language use in the Datavoz questionnaire was up-to-date and appropriate, and to better interpret the answers to some of the key questions. Third, as discussed in Section C.3, Prolific participants were asked additional questions after the list experiments to test whether they trusted the list experiment methodology to protect their privacy, whether they thought the instructions were clear, and whether they had guessed the topic of the study. In addition, in order to verify that this study actually measure taste-based rather than statistical discrimination, Prolific participants were asked to explain and motivate their answers to all three direct questions in order to test whether individuals expressing comfort or discomfort with gay men was driven by prejudice or preferences rather than by productivity expectations or beliefs (Section C.4).

Prolific has been used in many economics studies (Zmigrod, Rentfrow, and Robbins 2018; Schild et al. 2019; Isler, Maule, and Starmer 2018; Oreffice and Quintana-Domeque 2021; Aksoy, Carpenter, and Sansone 2023; 2025). Available evidence indicates some important advantages of Prolific over Amazon Mechanical Turk for conducting research: Prolific participants are more diverse, less dishonest, pay more attention to study instructions, and produce higher quality data (Peer et al. 2017; Palan and Schitter 2018; Peer et al. 2021; Gupta, Rigotti, and Wilson 2021).

This pilot collected responses from 535 participants from Chile between 18 July and 31 July 2024. The main study took about 11 minutes and 45 seconds on average to complete, and subjects who successfully completed the study received USD 3 which, on average, corresponds to USD 15.3/hour.

**C.1 Effect of list experiments on direct question.**

One may worry that participants' responses to the key direct questions may have been affected by their answers to the list experiments. For instance, it is possible that participants wanted to be consistent with their answers to the list experiments and report not being comfortable supervising with a gay employee if this was in line with the figure they reported after the related list experiment, even if they may have reported feeling comfortable supervising a gay employee if they had not seen the list experiment beforehand and they had felt social pressure to answer affirmatively.

This channel is somewhat unlikely since respondents saw a total of 24 items in the list experiments, and they were asked the related direct questions only about three of these items. Moreover, instead of asking the direct questions right after their corresponding lists, as done for instance in (Coffman,

Coffman, and Ericson 2017), the direct questions were located after the demographic questions, and together with other questions on income, religiosity, and political affiliation, in line with (Lax, Phillips, and Stollwerk 2016) and (Chuang et al. 2021). This choice should reduce the probability of respondents linking the direct questions to the related list experiment.

In addition, while (Lax, Phillips, and Stollwerk 2016) also asked the direct questions after their list experiment on support for same-sex marriage, they used a single experiment method which made it possible for them to study the impact of seeing the key item in the list experiment on direct survey question responses. Reassuringly, they did not find any significant impact coming from the fact that half of their subjects saw the key statement twice (once in the list experiment and once as a direct question).

To produce a similar test, only half of the respondents in the Prolific pilot were asked to respond to both the list experiments and the survey questions, while the other half were shown only the survey questions (including the direct questions). As shown in Table C1, average levels of comfort with gay individuals as reported in the direct questions is not affected by whether the respondents were also asked to participate in the list experiments before answering the direct questions.

**C.2 Terminology.**

The pilot on Prolific was also leveraged to make sure that the language use in the main questionnaire was up-to-date and appropriate. Indeed, respondents were asked which term they thought was more commonly used to refer to a person working with them. As shown in Table C2, "co-worker" was the most selected term. Therefore, this terminology was adopted in the main Datavoz survey.

In addition, respondents were asked which term they thought was more commonly used to refer to people who are attracted to individuals of the same sex. As shown in Table C3, gay was the most selected term. Therefore, this terminology was adopted in the main Datavoz survey.

Respondents were then asked whether they think that one is talking about gay men or lesbian women when one refers to gay or homosexual individuals. As shown in Table C4, most respondents thought that gay/homosexual referred mainly to both gay men and lesbian women. This finding is important to interpret from a gender perspective the main findings from the list experiments

**C.3 Feedback on survey instructions and survey aim.**

As in the main survey, participants in the Prolific pilot were also administered questions to provide feedback regarding the instructions and the instrument overall. Table C5 tabulates the responses to the prompt: "I believe the instructions were clear". All participants, except 4, agreed with the prompt. This question was presented to all participants at the end of the survey.

The subset of participants in the Prolific pilot who were randomized into the list experiment were also provided with questions to leave feedback for the researchers immediately after answering the list experiment and before continuing with the survey. Specifically, we used an open-ended question to ask respondents what they believed the study was about. Among the top 30 most frequent words in response to this question were: "social issues", "opinions", "people", "Chileans", "population", "politics", "beliefs", "ideas", and others. LGBTQ-related terms[10] were mentioned by only 9 out of 266 participants. Additionally, a manual inspection of all the non-empty open-ended responses confirms that most individuals are not sure or are confused about the purpose of the study, express curiosity about the goals of the researchers, and provide guesses along the lines of "eliciting beliefs and opinions from Chileans about controversial social and political issues". These findings confirm that participants did not believe that the list experiments were focused on measuring attitudes towards sexual minorities, thus achieving the goal of not priming or influencing respondents.

**C.4 Follow-up questions after the direct questions.**

Our study focuses on measuring sources of taste-based discrimination. A key goal of the Prolific pilot was to help us understand whether in fact taste-based preferences (as opposed to statistical discrimination) explained respondents' behaviors. To this end, the direct questions (i.e., "Would you feel comfortable supervising a gay employee?", "Would you feel comfortable working closely with a gay co-worker.", and "Would you feel comfortable having a cashier at the supermarket who is gay?") were immediately followed by mandatory open-ended text questions asking participants to explain their previous Yes/No responses.

To analyze these data, we created an indicator variable that flags respondents who justified their responses to any of the three statements with mentions of "productivity", "performance", "output", "efficacy", "efficiency", "stereotype", and "results". All the flagged responses expressed indifference on these qualifiers (i.e., that gay people do not differ from others on these considerations). Restricting the sample of open-ended explanations to the small subsets of respondents who answered negatively to the any of the direct questions reveals that most justifications are taste-based. For example, some participants express discomfort with the idea of a gay employee or coworker insinuating attraction towards them. Other participants mention that gay people contradict the conservative values their families raised them on. Only one participant mentioned a stereotype consisting of gay people being "conflictive" and "gossiping": the respondent argues this presents a risk at work due to the "current laws and political system". Overall, these findings confirm that the list experiments are measuring potential sources of taste-based discrimination, not statistical discrimination.

---

[10] LGBTQ-related terms that we searched for include: "lgbt", "lgbtq", "lgbtq+", "LGBT", "LGBTQ", "LGBTQ+", "gay", "lesbiana", "transgénero", "bisexual", "queer", "homosexual", "transexual", "género", "orientación sexual", and "diversidad sexual".

**Table C1: effect of list experiments on direct questions.**

|  | With list experiments | Direct questions only | Difference |
|---|---|---|---|
| Would you feel comfortable… | (1) | (2) | (2) – (1) |
| supervising a gay employee? | 0.96 | 0.97 | 0.01 |
|  | (0.20) | (0.18) | [0.6191] |
| working closely with a gay colleague? | 0.95 | 0.96 | 0.01 |
|  | (0.21) | (0.20) | [0.8032] |
| having a gay cashier at the supermarket? | 0.99 | 0.98 | -0.01 |
|  | (0.09) | (0.15) | [0.1610] |

Note: Column 1 reports the average responses to the three direct questions among participants who also responded to the list experiments. Column 2 reports the average responses to the three direct questions among participants who were not shown the list experiments. Column 3 reports the difference between the two groups. Standard deviation reported in parenthesis for Columns (1) and (2). P-value reported in square brackets for Column (3). *** p<0.01, ** p<0.05, * p<0.1

**Table C2: colleague versus co-worker.**

| Which term do you think is most common to refer to a person who works with you? | Number of observations | Share of respondents |
|---|---|---|
| Colega | 181 | 33.83 |
| Compañero de trabajo | 329 | 61.50 |
| Don't know / not sure | 25 | 4.67 |

Note: the original Spanish question is "¿Qué término crees que es más común para referirse a una persona que trabaja contigo?".

**Table C3: gay versus homosexual.**

| Which term do you think is most common to refer to a person who is attracted to people of the same sex? | Number of observations | Share of respondents |
|---|---|---|
| Gay | 356 | 66.54 |
| Homosexual | 142 | 26.54 |
| Don't know / not sure | 37 | 6.92 |

Note: the original Spanish question is "¿Qué término crees que es más común para referirse a una persona que siente atracción hacia personas del mismo sexo?".

**Table C4: gay/homosexual and gender.**

| When we refer to gay/homosexual people, do you think we mean...? | Number of observations | Share of respondents |
|---|---|---|
| Both | 398 | 74.39 |
| Gay/homosexual men | 132 | 24.67 |
| Gay/homosexual/lesbian women | 1 | 0.19 |
| Don't know / not sure | 4 | 0.75 |

Note: the original Spanish question is "Cuando nos referimos a personas gays/homosexuales, ¿crees que nos referimos a...?".

**Table C5. Instructions feedback – All participants**

| Indicate the extent to which you agree or disagree with the following statement: "The instructions were clear" | Number of observations | Share of respondents |
|---|---|---|
| Totally agree | 517 | 96.64 |
| Somewhat agree | 14 | 2.62 |
| Neither agree nor disagree | 0 | 0.19 |
| Somewhat disagree | 1 | 0 |
| Totally disagree | 3 | 0.56 |

Note: the original Spanish question is "Indique en qué medida está de acuerdo o en desacuerdo con la siguiente afirmación: "Las instrucciones fueron claras"."

**Appendix D. Additional tables and figures.**

**Figure D1. List Experiment on Attitudes from Gay Supervisors, Co-workers, and Customers - Heterogeneity by Sex Assigned at Birth**

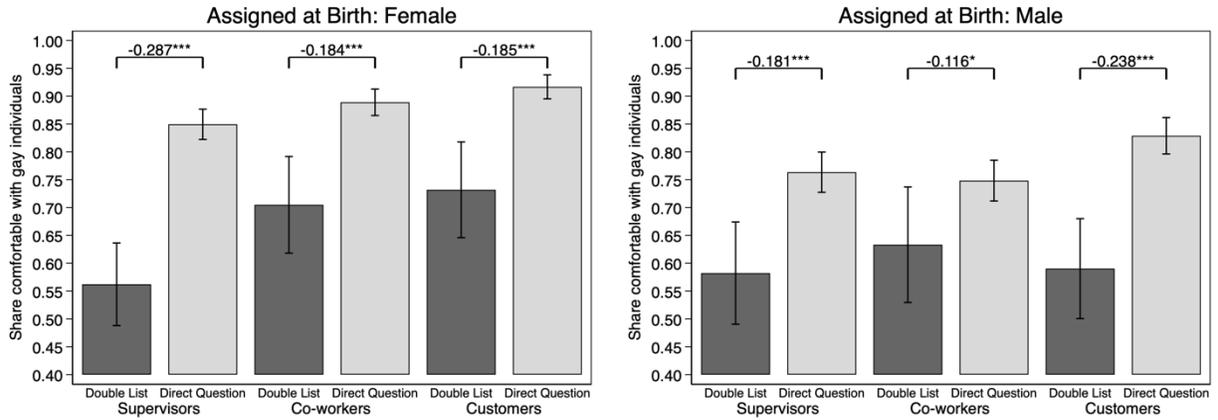

*Notes:* Weighted statistics. 95% confidence intervals are reported with the vertical range plots. The numbers above the horizontal bars are the differences between the two groups at the base of each horizontal bar. Supervisor key statement: "I would feel comfortable supervising a gay employee." Co-workers key statement: "I would feel comfortable working closely with a gay colleague." Customers key statement: "I would feel comfortable having a gay cashier at the supermarket." Number of observations: 1,879 (Assigned Female at Birth) and 2,121 (Assigned Male at Birth). *p < 0.10; **p < 0.05; ***p < 0.01.

**Figure D2. List Experiment on Attitudes from Gay Supervisors, Co-workers, and Customers - Heterogeneity by Sexual Orientation**

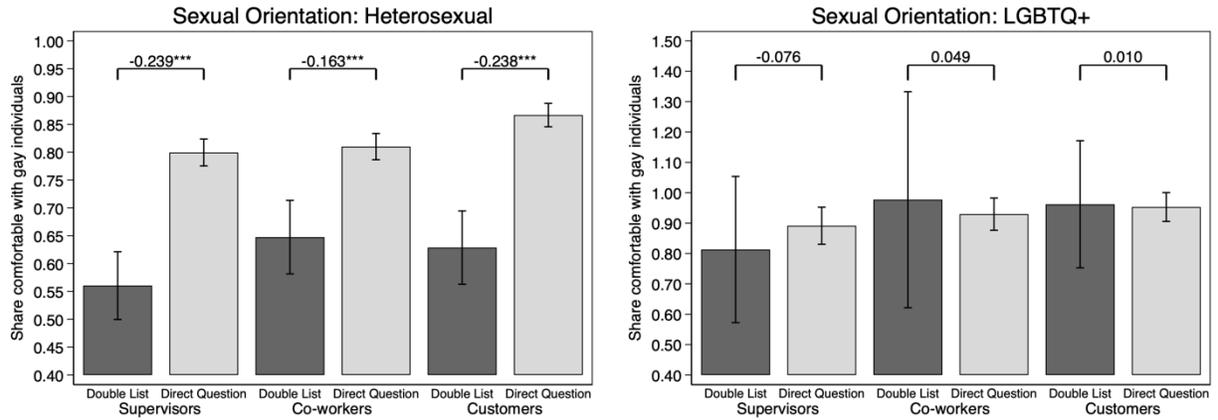

*Notes:* Weighted statistics. 95% confidence intervals are reported with the vertical range plots. The numbers above the horizontal bars are the differences between the two groups at the base of each horizontal bar. Supervisor key statement: "I would feel comfortable supervising a gay employee." Co-workers key statement: "I would feel comfortable working closely with a gay colleague." Customers key statement: "I would feel comfortable having a gay cashier at the supermarket." Number of observations: 3,659 (Heterosexual) and 293 (LGBTQ+). *p < 0.10; **p < 0.05; ***p < 0.01.

**Figure D3. List Experiment on Attitudes from Gay Supervisors, Co-workers, and Customers - Heterogeneity by Managerial Experience**

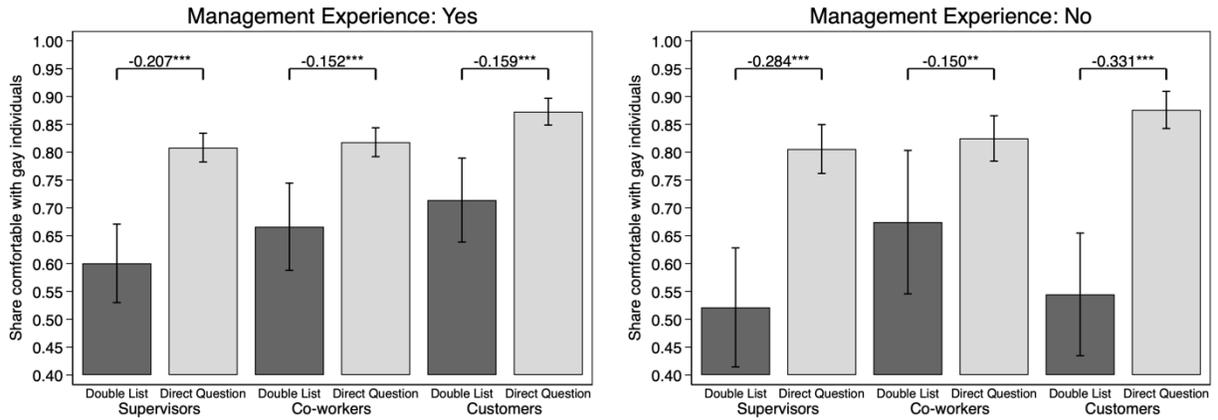

*Notes:* Weighted statistics. 95% confidence intervals are reported with the vertical range plots. The numbers above the horizontal bars are the differences between the two groups at the base of each horizontal bar. Supervisor key statement: "I would feel comfortable supervising a gay employee." Co-workers key statement: "I would feel comfortable working closely with a gay colleague." Customers key statement: "I would feel comfortable having a gay cashier at the supermarket." Number of observations: 3,076 (Management Experience: Yes) and 924 (Management Experience: No). *p < 0.10; **p < 0.05; ***p < 0.01.

**Figure D4. List Experiment on Attitudes from Gay Supervisors, Co-workers, and Customers - Heterogeneity by Religiosity**

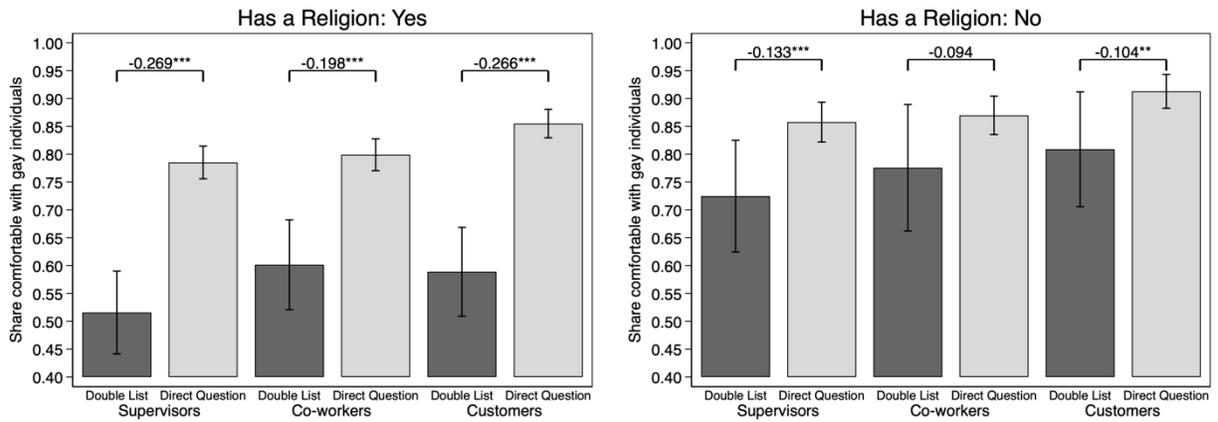

*Notes:* Weighted statistics. 95% confidence intervals are reported with the vertical range plots. The numbers above the horizontal bars are the differences between the two groups at the base of each horizontal bar. Supervisor key statement: "I would feel comfortable supervising a gay employee." Co-workers key statement: "I would feel comfortable working closely with a gay colleague." Customers key statement: "I would feel comfortable having a gay cashier at the supermarket." Number of observations: 2,456 (Has a Religion: Yes) and 1,443 (Has a Religion: No). *p < 0.10; **p < 0.05; ***p < 0.01.

**Figure D5. List Experiment on Attitudes from Gay Supervisors, Co-workers, and Customers - Heterogeneity by Political Affiliation**

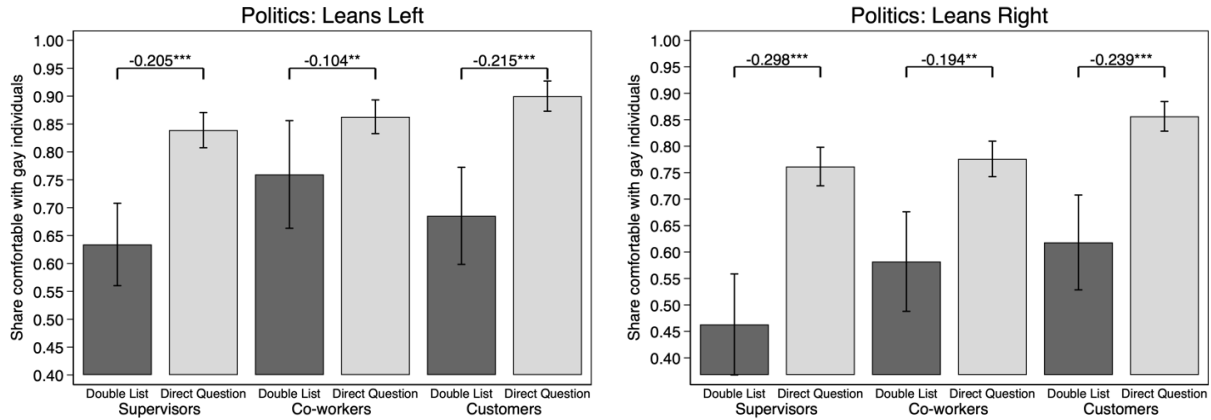

*Notes:* Weighted statistics. 95% confidence intervals are reported with the vertical range plots. The numbers above the horizontal bars are the differences between the two groups at the base of each horizontal bar. Supervisor key statement: "I would feel comfortable supervising a gay employee." Co-workers key statement: "I would feel comfortable working closely with a gay colleague." Customers key statement: "I would feel comfortable having a gay cashier at the supermarket." Number of observations: 2,043 (Politics: Leans Left) and 1,720 (Politics: Leans Right). *p < 0.10; **p < 0.05; ***p < 0.01.

**Figure D6. List Experiment on Attitudes from Gay Supervisors, Co-workers, and Customers - Heterogeneity by Education**

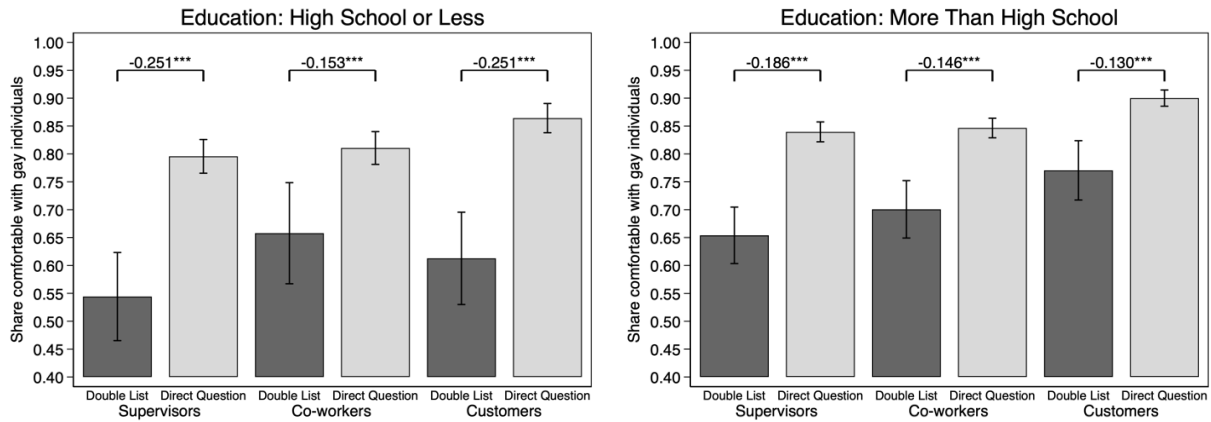

*Notes:* Weighted statistics. 95% confidence intervals are reported with the vertical range plots. The numbers above the horizontal bars are the differences between the two groups at the base of each horizontal bar. Supervisor key statement: "I would feel comfortable supervising a gay employee." Co-workers key statement: "I would feel comfortable working closely with a gay colleague." Customers key statement: "I would feel comfortable having a gay cashier at the supermarket." Number of observations: 1,197 (Education: high School or Less) and 2,803 (Education: More Than High School). *$p < 0.10$; **$p < 0.05$; ***$p < 0.01$.

**Figure D7. List Experiment on Attitudes from Gay Supervisors, Co-workers, and Customers - Heterogeneity by LGB Familiarity**

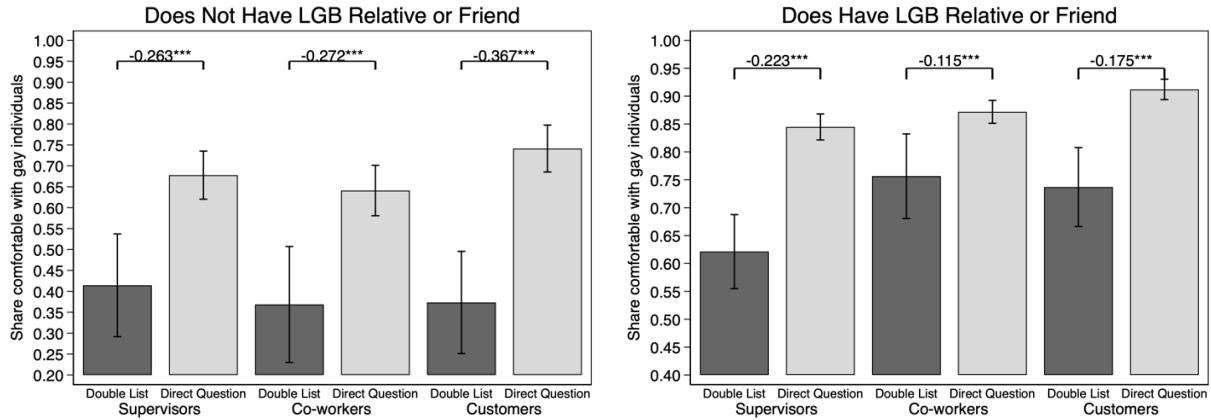

*Notes:* Weighted statistics. 95% confidence intervals are reported with the vertical range plots. The numbers above the horizontal bars are the differences between the two groups at the base of each horizontal bar. Supervisor key statement: "I would feel comfortable supervising a gay employee." Co-workers key statement: "I would feel comfortable working closely with a gay colleague." Customers key statement: "I would feel comfortable having a gay cashier at the supermarket." Number of observations: 793 (Does Not Have LGB Relative or Friend) and 3,207 (Does Have LGB Relative or Friend). *p < 0.10; **p < 0.05; ***p < 0.01.

**Table D1. List Experiment on Attitudes Toward Gay Individuals by Supervisors, Co-workers and Customers – Heterogeneity Analyses**

| Interaction of treatment variable with: | Supervisors (1) | Co-workers (2) | Customers (3) |
|---|---|---|---|
| Age: 18–44 | 0.102 | 0.099 | -0.066 |
|  | (0.065) | (0.068) | (0.064) |
| Race: African Descent | -0.323 | 0.012 | 0.086 |
|  | (0.200) | (0.173) | (0.196) |
| Indigenous | -0.005 | 0.108 | -0.063 |
|  | (0.110) | (0.122) | (0.123) |
| Sex assigned at birth: Female | -0.035 | 0.095 | 0.126* |
|  | (0.061) | (0.069) | (0.066) |
| Sexual orientation: Heterosexual | -0.149 | -0.198 | -0.306*** |
|  | (0.114) | (0.174) | (0.115) |
| Household income: More than $975 | 0.063 | -0.183*** | -0.021 |
|  | (0.058) | (0.063) | (0.065) |
| Education: Less than high school | -0.016 | -0.054 | -0.125 |
|  | (0.078) | (0.086) | (0.079) |
| Employment status: Employed | 0.084 | 0.112 | -0.047 |
|  | (0.080) | (0.075) | (0.082) |
| Management Experience | 0.118* | 0.098 | 0.142** |
|  | (0.069) | (0.077) | (0.067) |
| Region: Outside metro | 0.017 | -0.089 | -0.042 |
|  | (0.057) | (0.062) | (0.059) |
| Political affiliation: Lean left | 0.074 | -0.034 | -0.029 |
|  | (0.067) | (0.074) | (0.067) |
| Current religious affiliation: Not religious | 0.106 | 0.084 | 0.207*** |
|  | (0.065) | (0.076) | (0.071) |
| LGBTQ+ Comfort | 0.391*** | 0.328*** | 0.252*** |
|  | (0.074) | (0.077) | (0.075) |
| Has Gay Relative(s) and/or Friend(s) | 0.012 | 0.242*** | 0.185** |
|  | (0.085) | (0.088) | (0.080) |
| Belief: 50% or more comfortable supervising gay employees | 0.066 |  |  |
|  | (0.060) |  |  |
| Belief: 50% or more comfortable with gay co-workers |  | 0.124* |  |
|  |  | (0.066) |  |
| Belief: 50% or more comfortable with gay cashiers |  |  | 0.013 |
|  |  |  | (0.068) |
| Constant | 0.131 | 0.259 | 0.539*** |
|  | (0.169) | (0.215) | (0.172) |
| Observations | 3,646 | 3,646 | 3,646 |

*Notes*: Heterogeneity analysis. Multivariate analysis. Robust standard errors are in parentheses. Coefficients obtained using the Stata command kict ls (Tsai 2019) performing least squares estimation for a double list experiment. The dependent variables are the reported number of true statements for the supervisor of a gay employee lists (Column 1), the gay co-workers lists (Column 2), and the gay cashier lists (Column 3). The treatment variable is an indicator equal to 1 for the first long list (List A) containing the corresponding key statement and the second short list (List B), and equal to 0 for the first short list (List A) and the second long list (List B). *$p < 0.10$; **$p < 0.05$; ***$p < 0.01$.

**Table D2. List Experiments. Differences-in-means comparisons and robustness checks**

|  | List A | List B | Double List | Direct Question | (1)-(2) | (4)-(3) |
|---|---|---|---|---|---|---|
|  | (1) | (2) | (3) | (4) | (5) | (6) |
| *Panel A: Main Data (Weighted)* | | | | | | |
| Supervisor | 0.676 | 0.472 | 0.574 | 0.807 | 0.204** | 0.234*** |
|  | (0.054) | (0.050) | (0.030) | (0.012) | (0.085) | (0.030) |
| Co-worker | 0.772 | 0.567 | 0.669 | 0.820 | 0.205** | 0.151*** |
|  | (0.054) | (0.046) | (0.034) | (0.011) | (0.073) | (0.034) |
| Customer | 0.685 | 0.628 | 0.657 | 0.874 | 0.057 | 0.217*** |
|  | (0.054) | (0.055) | (0.032) | (0.010) | (0.088) | (0.032) |
| *Panel B: Main Data (Unweighted)* | | | | | | |
| Supervisor | 0.649 | 0.604 | 0.627 | 0.827 | 0.045 | 0.200*** |
|  | (0.024) | (0.025) | (0.016) | (0.006) | (0.036) | (0.016) |
| Co-worker | 0.693 | 0.644 | 0.668 | 0.838 | 0.049 | 0.170*** |
|  | (0.025) | (0.024) | (0.016) | (0.006) | (0.036) | (0.016) |
| Customer | 0.696 | 0.761 | 0.728 | 0.893 | -0.065 | 0.165*** |
|  | (0.025) | (0.027) | (0.017) | (0.005) | (0.038) | (0.017) |
| *Panel C: Main Data and Pilots (Unweighted)* | | | | | | |
| Supervisor | 0.621 | 0.581 | 0.621 | 0.828 | 0.040 | 0.206*** |
|  | (0.016) | (0.016) | (0.016) | (0.006) | (0.034) | (0.016) |
| Co-worker | 0.669 | 0.623 | 0.669 | 0.837 | 0.046 | 0.168*** |
|  | (0.016) | (0.016) | (0.016) | (0.006) | (0.034) | (0.016) |
| Customer | 0.723 | 0.670 | 0.723 | 0.893 | -0.053 | 0.170*** |
|  | (0.017) | (0.037) | (0.032) | (0.005) | (0.037) | (0.017) |

*Notes*: Standard errors are in parentheses. Columns (1) and (2) report the mean differences in responses across treatments in List A and List B, respectively. Column (3) reports the share of individuals comfortable with the corresponding key statement elicited by the double list experiment. Column (4) reports the mean of the direct question for each corresponding statement. Column (5) reports differences between Column (1) and (2) and Column (6) reports differences between Column (4) and (3). Supervisor key statement: "I would feel comfortable supervising a gay employee." Co-workers key statement: "I would feel comfortable working closely with a gay colleague." Customers key statement: "I would feel comfortable having a gay cashier at the supermarket." Number of observations: 4,000 (Panels A and B), and 4,297 (Panel C). *p < 0.10; **p < 0.05; ***p < 0.01.

**Appendix E. Datavoz questionnaire.**

**E.1 Recruitment emails**

**E.1.1 Original Spanish Version of Recruitment Email**

**ASUNTO**: Invitación a participar en encuesta BID-DATAVOZ

> Datavoz por encargo del Banco Interamericano del Desarrollo (BID) se encuentra realizando una encuesta. La encuesta implica responder a preguntas bajo diferentes escenarios según sus preferencias individuales.
>
> La participación debería tomar aproximadamente 10 minutos.
>
> **Se sortearán 50 tarjetas de regalo (gift cards) de $50.000 pesos chilenos entre los participantes que completen la encuesta. Además, un subconjunto aleatorio de 50 encuestados podrá recibir $100.000 pesos chilenos adicionales en forma de gift cards.**
>
> **<<ENLACE DE ENCUESTA>>**
>
> Muchas gracias

### E.1.2 English Version of Recruitment Email

**SUBJECT LINE**: Invitation to participate in survey IDB-DATAVOZ

Datavoz, commissioned by the Inter-American Development Bank (IDB), is conducting a survey. The survey involves answering questions under different scenarios based on your individual preferences.

Participation should take approximately 10 minutes.

**Fifty gift cards worth 50,000 Chilean pesos each will be raffled among the participants who complete the survey. Additionally, a random subset of 50 respondents may receive an extra 100,000 Chilean pesos in the form of gift cards.**

<<**LINK TO SURVEY**>>

Thank you very much.

## E.2 Consent Forms.

### E.2.1 Original Spanish Version of Consent Form.

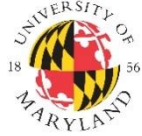

**Institutional Review Board**
1204 Marie Mount Hall ● 7814 Regents Drive ● College Park, MD 20742 ● 301-405-4212 ● irb@umd.edu

**CONSENTIMIENTO PARA PARTICIPAR**

| Título del Proyecto | Encuesta de Opiniones sobre Cuestiones Sociales en Chile |
|---|---|
| Investigador Principal | Esta investigación es realizada por Ariel Listo de la Universidad de Maryland, College Park, Ercio Munoz del Banco Interamericano de Desarrollo (BID), y Dario Sansone de la Universidad de Exeter, en colaboración con DATAVOZ. |
| Propósito del Estudio | El propósito de este proyecto de investigación es entender las opiniones de las personas sobre ciertas cuestiones sociales. |
| Procedimientos | La encuesta implica responder a preguntas bajo diferentes escenarios según sus preferencias individuales. En algunos casos, sus decisiones pueden tener un efecto real en sus ganancias. La participación debería tomar aproximadamente 10 minutos. |
| Potenciales Riesgos y Molestias | No existen riesgos físicos, psicológicos o sociales conocidos para los participantes aparte de la ansiedad o el aburrimiento asociados con la participación en este proyecto de investigación. Puede tomar descansos según sea necesario. |
| Beneficios Potenciales | Esta investigación no está diseñada para ayudarlo personalmente, pero los resultados pueden ayudar a los investigadores a aprender más sobre las opiniones sociales. Esperamos que en el futuro otras personas puedan beneficiarse de este estudio a través de una mejor comprensión de las preferencias individuales. |
| Confidencialidad | No accederemos a ninguna información personal identificable sobre usted. Los investigadores no recibirán datos identificadores sobre usted. Todos los datos no identificables se almacenarán en computadoras protegidas por contraseña y se compartirán solo a través de servicios en línea seguros.<br><br>Si escribimos un informe o artículo sobre este proyecto de investigación, su identidad será protegida al máximo posible. Su información puede ser compartida con representantes de la Universidad de Maryland, College Park o autoridades gubernamentales si usted o alguien más está en peligro o si estamos obligados a hacerlo por ley. |
| Compensación | Se sortearán 50 tarjetas de regalo (gift cards) de $50.000 pesos chilenos entre los participantes que completen la encuesta. Además, un subconjunto aleatorio de 50 encuestados podrán recibir $100.000 pesos chilenos adicionales en forma de gift cards.<br><br>Si decide no terminar la encuesta, no entrará en la rifa ni recibirá ningún tipo de compensación parcial. Usted será responsable de cualquier impuesto evaluado sobre la compensación. |
| **Derecho a Retirarse y Preguntas** | Su participación en esta investigación es completamente voluntaria. Puede optar por no participar en absoluto. Si decide participar en esta investigación, puede dejar de participar en cualquier momento. Si decide no participar en este estudio o si deja de participar en cualquier momento, no será penalizado ni perderá ningún beneficio para el cual de otra manera calificaría, pero no entrará en la rifa ni recibirá ningún tipo de compensación parcial.<br><br>Si decide dejar de participar en el estudio, si tiene preguntas, inquietudes o quejas, o si necesita informar una lesión relacionada con la investigación, por favor contacte al investigador:<br>Ariel Listo<br>2200 Symons Hall, University of Maryland, College Park, MD 20742 USA<br>alisto@umd.edu<br>(301) 405-1293<br><br>Dr. Dario Sansone<br>University of Exeter Business School, Rennes Drive, Exeter EX4 4PU, Reino Unido<br>Email: d.sansone@exeter.ac.uk |
| **Derechos del Participante** | Si tiene preguntas sobre sus derechos como participante en la investigación o desea informar una lesión relacionada con la investigación, por favor contacte:<br><br>University of Maryland College Park<br>Institutional Review Board Office<br>1204 Marie Mount Hall<br>College Park, Maryland, 20742 USA<br>E-mail: irb@umd.edu<br>Telephone: 301-405-0678<br><br>Para más información sobre los derechos de los participantes, por favor visite: https://research.umd.edu/research-resources/research-compliance/institutional-review-board-irb/research-participants<br><br>Esta investigación ha sido revisada de acuerdo con los procedimientos de IRB de la Universidad de Maryland, College Park para investigaciones que involucran a sujetos humanos. |
| **Declaración de Consentimiento** | Su consentimiento indica que usted tiene al menos 18 años de edad; ha leído este formulario de consentimiento o se lo han leído; y acepta voluntariamente participar en este estudio de investigación. Puede descargar esta hoja informativa para sus registros. |

**E.2.1 English Translation of Consent Form**

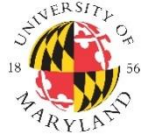

**Institutional Review Board**
1204 Marie Mount Hall ● 7814 Regents Drive ● College Park, MD 20742 ● 301-405-4212 ● irb@umd.edu

## CONSENT TO PARTICIPATE

| | |
|---|---|
| **Project Title** | Survey on Opinions about Social Issues in Chile |
| **Principal Investigator** | This research is being conducted by Ariel Listo at the University of Maryland, College Park, Ercio Munoz at the Inter-American Development Bank (IDB), and Dario Sansone at the University of Exeter, in collaboration with DATAVOZ. |
| **Purpose of the Study** | The purpose of this research project is to understand individuals' views on certain social issues. |
| **Procedures** | The survey involves responding to questions and making choices under different scenarios according to your individual preferences. In some cases, your decisions may have a real effect on your earnings. Participation should take about 10 minutes. |
| **Potential Risks and Discomforts** | There are no known physical, psychological, or social risks to subjects, other than anxiety or boredom associated with participating in this research project. You may take breaks as needed. |
| **Potential Benefits** | This research is not designed to help you personally, but the results may help the investigators learn more about social views. We hope that, in the future, other people might benefit from this study through improved understanding of individuals' preferences. |
| **Confidentiality** | We will not be accessing any personally identifying information about you. No identifiers will be shared with the investigators. All non-identifiable data will be stored on password-protected computers and shared only through secure online services.<br><br>If we write a report or article about this research project, your identity will be protected to the maximum extent possible. Your information may be shared with representatives of the University of Maryland, College Park or governmental authorities if you or someone else is in danger or if we are required to do so by law. |
| **Compensation** | 50 gift cards of 50,000 Chilean pesos each will be raffled among the participants who complete the survey. Additionally, a random subset of 50 respondents may receive an additional 100,000 Chilean pesos.<br><br>Should you choose to not finish the survey, you will not be entered into the raffle or receive any type of partial compensation. You will be responsible for any taxes assessed on the compensation. |
| **Right to Withdraw and Questions** | Your participation in this research is completely voluntary. You may choose not to take part at all. If you decide to participate in this research, you may stop participating at any time. If you decide not to participate in this study or if you stop participating at any time, you will not be penalized or lose any benefits to which you |

|  | otherwise qualify, but you will not be entered into the raffle or receive any type of partial compensation. |
|---|---|
|  | If you decide to stop taking part in the study, if you have questions, concerns, or complaints, or if you need to report an injury related to the research, please contact the investigators:<br><br>Ariel Listo<br>2200 Symons Hall, University of Maryland, College Park, MD 20742 USA<br>Email: alisto@umd.edu<br>(301) 405-1293<br><br>Dr. Dario Sansone<br>University of Exeter Business School, Rennes Drive, Exeter EX4 4PU, UK<br>Email: d.sansone@exeter.ac.uk |
| **Participant Rights** | If you have questions about your rights as a research participant or wish to report a research-related injury, please contact:<br><br>University of Maryland College Park<br>Institutional Review Board Office<br>1204 Marie Mount Hall<br>College Park, Maryland, 20742 USA<br>E-mail: irb@umd.edu<br>Telephone: 301-405-0678<br><br>For more information regarding participant rights, please visit:<br>https://research.umd.edu/research-resources/research-compliance/institutional-review-board-irb/research-participants<br><br>This research has been reviewed according to the University of Maryland, College Park IRB procedures for research involving human subjects. |
| **Statement of Consent** | Your consent indicates that you are at least 18 years of age; you have read this consent form or have had it read to you; and you voluntarily agree to participate in this research study. You may download this information sheet for your records. |

## E.3 Questionnaire.

¡Bienvenido!

Gracias por elegir participar de nuestra encuesta.

Antes de continuar, por favor completa la siguiente verificación.

☐ No soy un robot  reCAPTCHA
Privacidad - Términos

SIGUIENTE

Por favor, revisa la información contenida en el formulario a continuación.

[Consentimiento Informado](#)

Haciendo clic en [Sí, consiento...] a continuación, confirmas que tienes al menos 18 años de edad y que consientes participar en el estudio de investigación descrito anteriormente.

Si haces clic en [No, no consiento...], no serás elegible para participar en este estudio y serás redirigido al final de la encuesta.

> Si, consiento a participar en este estudio.

> No, no consiento a participar en este estudio.

**SIGUIENTE**

## Encuesta de Opiniones sobre Cuestiones Sociales en Chile

**Introducción**

En este estudio, se te pedirá que respondas algunas preguntas. Tus respuestas serán anónimas. El estudio tomará aproximadamente **10 minutos.**

**¿Cuánto recibiré de pago por mi participación?**

<u>Sorteo por participación:</u> Se sortearán 50 gift cards de $50.000 pesos entre los participantes que completen la encuesta. Este sorteo se realizará luego de que todos los participantes de esta encuesta, finalicen su participación.

<u>Sorteo por bono adicional:</u> Además del sorteo por participación, se realizará un sorteo antes que tu termines la encuesta en la cual un subconjunto aleatorio de 50 encuestados podrán recibir $100.000 pesos adicionales en forma de gift cards.

DATAVOZ utilizará el correo electrónico a través del cual te contactamos para hacerte llegar tu premio si resultaras ganador de uno de estos sorteos.

**Preguntas de Verificación de Atención**

En cuestionarios como este, a veces hay participantes que no leen cuidadosamente las preguntas y solo hacen clic rápidamente a través de la encuesta. Esto puede comprometer los resultados de estudios de investigación como este. Como prevención, habrá varias preguntas de **Verificación de Atención** a lo largo de esta encuesta. Si no logras completar correctamente dos o más de estas preguntas, es posible que no recibas compensación.

Por favor, haz clic en SIGUIENTE cuando estés listo.

**SIGUIENTE**

## Instrucciones

En las siguientes páginas, se presentarán listas de afirmaciones que pueden ser o no verdaderas para ti. Las afirmaciones serán sobre ti mismo y tus opiniones sobre temas sociales. Nos gustaría saber cuántas de las afirmaciones en cada lista son verdaderas para ti. En estas listas, no estamos preguntando **cuáles** afirmaciones específicas son verdaderas para ti, solo estamos preguntando **cuántas** de ellas son verdaderas para ti. No existen respuestas buenas o malas. Solo nos interesa conocer su opinión y experiencia.

En la página siguiente, te daremos un ejemplo.

Por favor, haz clic en SIGUIENTE cuando estés listo.

SIGUIENTE

**Ejemplo**

Queremos asegurarnos de que entiendas cómo funciona esto. Aquí tienes un ejemplo:

\*\*\*

Hay tres (3) afirmaciones en la siguiente lista. ¿Cuántas de estas afirmaciones son verdaderas para ti?

- Tengo una camiseta naranja.
- En mi hogar hay al menos cuatro mascotas.
- Reciclo regularmente

Por favor, ingresa el número total de estas afirmaciones que son verdaderas para ti: \_\_\_\_

\*\*\*

*Supongamos que sí tienes una camiseta naranja y que en tu hogar hay al menos cuatro mascotas. Pero no reciclas. En ese caso, dos de las afirmaciones anteriores son verdaderas para ti. Por lo tanto, indicarías esto ingresando 2 en el cuadro de respuesta.*

**Ten en cuenta:** No estamos preguntando **cuáles** afirmaciones específicas de estas listas son verdaderas para ti. Solo estamos preguntando **cuántas** de ellas son verdaderas para ti.

Por favor, haz clic en SIGUIENTE cuando estés listo.

SIGUIENTE

Si deseas revisar las instrucciones una vez más, por favor haz clic aquí.

Cuando estés listo, por favor haz clic en SIGUIENTE para comenzar.

SIGUIENTE

Hay cinco (5) afirmaciones en la siguiente lista. ¿Cuántas de estas afirmaciones son verdaderas para ti?

- Tengo una conexión a internet en mi casa.
- Estoy a favor de la legalización del uso recreativo de la marihuana.
- He conocido personalmente a Arturo Vidal.
- El gobierno debe reducir los impuestos a las empresas.
- Me sentiría cómodo trabajando estrechamente con un compañero de trabajo gay.

Por favor, ingresa el número total de estas afirmaciones que son verdaderas para ti:

SIGUIENTE

Hay cuatro (4) afirmaciones en la siguiente lista. ¿Cuántas de estas afirmaciones son verdaderas para ti?

- Estoy de acuerdo con que Chile acepte a más refugiados de otros países.
- Tengo un computador.
- He conocido personalmente a Don Francisco (Mario Kreutzberger).
- Estoy de acuerdo con la prohibición del aborto.

Por favor, ingresa el número total de estas afirmaciones que son verdaderas para ti:

[ ]

SIGUIENTE

Hay cinco (5) afirmaciones en la siguiente lista. ¿Cuántas de estas afirmaciones son verdaderas para ti?

- Estoy en contra de las protestas que obstruyen el paso de los vehículos.
- Me sentiría cómodo siendo atendido en el supermercado por un cajero que es gay.
- Creo que en Chile no hay políticos corruptos en el Congreso.
- Creo que en todas las escuelas se debería enseñar educación sexual.
- Tengo una licencia de conducir.

Por favor, ingresa el número total de estas afirmaciones que son verdaderas para ti:

[ ]

SIGUIENTE

Hay cuatro (4) afirmaciones en la siguiente lista. ¿Cuántas de estas afirmaciones son verdaderas para ti?

- Creo que la mitad de los legisladores en el Congreso deben ser mujeres.
- Conozco a alguien que tiene un auto o motocicleta.
- Creo que los militares deben trabajar con la policía para combatir el crimen.
- Tengo mucha confianza en los partidos políticos.

Por favor, ingresa el número total de estas afirmaciones que son verdaderas para ti:

SIGUIENTE

Hay cinco (5) afirmaciones en la siguiente lista. ¿Cuántas de estas afirmaciones son verdaderas para ti?

- He visitado más de veinte países.
- Creo que las leyes de protección del medio ambiente no son lo suficientemente estrictas para combatir el cambio climático.
- Creo que los pobres hacen pocos esfuerzos para salir de la pobreza.
- Tengo, por lo menos, una cuenta de redes sociales (por ejemplo, Facebook, TikTok, Instagram, u otras).
- Me sentiría cómodo supervisando a un empleado gay.

Por favor, ingresa el número total de estas afirmaciones que son verdaderas para ti:

SIGUIENTE

Hay cuatro (4) afirmaciones en la siguiente lista. ¿Cuántas de estas afirmaciones son verdaderas para ti?

- Creo que está mal aplicar la pena de muerte, sin importar el delito.
- Puedo hablar al menos tres idiomas fluidamente.
- Tengo un teléfono móvil.
- Creo que las mujeres deben ser las responsables del cuidado de los niños.

Por favor, ingresa el número total de estas afirmaciones que son verdaderas para ti:

[                    ]

SIGUIENTE

Hay cinco (5) afirmaciones en la siguiente lista. ¿Cuántas de estas afirmaciones son verdaderas para ti?

- Normalmente respondo a mis correos electrónicos en menos de 24 horas.
- Me preocupa que los medios de comunicación en mi país estén sesgados.
- Por favor, ingresa 7 como tu respuesta abajo, independientemente de cuantas de las otras afirmaciones sean verdaderas para ti.
- Esto es porque nos gustaría comprobar que estas leyendo cada ítem cuidadosamente.
- Nuevamente, por favor, ingresa 7 como tu respuesta abajo.

Por favor, ingresa el número total de estas afirmaciones que son verdaderas para ti:

[          ]

SIGUIENTE

A continuación, te haremos algunas preguntas demográficas sobre ti, así como tu opinión sobre ciertos temas. Por favor, responde las siguientes preguntas lo mejor que puedas. Nuevamente, recuerda que tus respuestas serán completamente anónimas.

SIGUIENTE

¿Cuántos años cumplidos tiene?

[          ]

¿Es o se considera perteneciente a algún pueblo indígena u originario?

- Si
- No

De acuerdo con sus antepasados, tradiciones y cultura, es o se considera (Recuerde que las personas afrodescendientes tienen antepasados africanos):

- Afrodescendiente
- Afrochileno/a
- Negro/a
- Del Pueblo Tribal Afrodescendiente Chileno
- Moreno/a de Azapa
- Negro/a de la Chimba
- Ninguna de las anteriores

SIGUIENTE

¿Cuál es su estado conyugal o civil actual?

- Casado/a
- Conviviente o pareja sin acuerdo de unión civil
- Conviviente civil (con acuerdo de unión civil)
- Anulado/a
- Separado/a
- Divorciado/a
- Viudo/a
- Soltero/a

¿Cuántas personas viven como miembros de su hogar, incluyendo a Ud. y a los niños, si los hay?

[____________]

¿Tiene usted hijos?

- Si
- No

SIGUIENTE

¿Cuál es su nivel educativo más alto alcanzado? (Una persona alcanzó o llegó a un nivel educativo cuando declara haber finalizado por lo menos un curso del nivel correspondiente)

- Nunca asistió
- Sala cuna
- Jardín infantil
- Prekínder
- Kínder
- Educación especial o preferencial
- Educación básica
- Primaria (sistema antiguo)
- Media científico humanista o artística
- Media técnico profesional
- Humanidades (Sistema antiguo)
- Técnico comercial, industrial, normalista (sistema antiguo)
- Técnico nivel superior (1 a 3 años) (incluye suboficial FFAA)
- Professional (4 años o más) (incluye oficial FFAA)
- Magister
- Doctorado

SIGUIENTE

La semana pasada (corresponde al período entre lunes y domingo anterior a la entrevista):

Trabajó...

- Trabajó por un pago en dinero o especies.
- Trabajó sin pago para un familiar.

No trabajó...

- Tenía empleo, pero estuvo de vacaciones, con licencia, en descanso laboral, etc.
- Se encontraba buscando empleo y disponible para trabajar.
- Estaba estudiando.
- Es jubilado/a, pensionado/a o rentista.
- Realizó quehaceres de su hogar.
- Otra situación (por favor especifique)

¿Ha tenido alguna experiencia laboral como supervisor de uno o más trabajadores?

- Si
- No

**SIGUIENTE**

¿En qué región vives?

¿En qué región o país naciste?

SIGUIENTE

Ahora, vamos a hacer preguntas que algunas personas podrían considerar sensibles. Como recordatorio, tus respuestas son anónimas.

SIGUIENTE

¿Te sentirías cómodo/a supervisando a un empleado gay?

- Si
- No

**SIGUIENTE**

¿Te sentirías cómodo/a trabajando estrechamente con un compañero de trabajo gay?

Si

No

SIGUIENTE

¿Te sentirías cómodo/a siendo atendido en el supermercado por un cajero que es gay?

Si

No

SIGUIENTE

¿Te sentirías cómodo/a...

| | Si | No |
|---|---|---|
| ...teniendo de vecino a una persona gay? | ○ | ○ |
| ...teniendo un dentista gay? | ○ | ○ |
| ...siendo atendido por un mesero gay? | ○ | ○ |
| ...tratando con un agente inmobiliario gay? | ○ | ○ |
| ...teniendo un jefe gay? | ○ | ○ |
| ...trabajando estrechamente con una compañera de trabajo lesbiana? | ○ | ○ |
| ...trabajando estrechamente con un compañero de trabajo perteneciente a un pueblo originario? | ○ | ○ |
| ...teniendo un conductor de taxi gay? | ○ | ○ |
| ...supervisando a varios trabajadores? | ○ | ○ |

SIGUIENTE

Si alguien que conocieras te revelara que es gay, ¿mantendrías la misma cercanía con esa persona?

- Si, mantendría la misma cercanía
- No, me distanciaría de esa persona
- No, me sentiría mas cercano
- No sé o no estoy seguro

SIGUIENTE

¿Entre sus familiares cercanos, parientes, vecinos, compañeros de trabajo o amigos íntimos, hay alguno que sea gay, lesbiana o bisexual (que usted sepa)?

- Si
- No

SIGUIENTE

Antes de dar una respuesta, siempre se debe leer el texto con atención. Para verificar si has leído el texto con atención, te pedimos que selecciones la tercera opción a continuación como tu respuesta

- Primera
- Segunda
- Tercera
- Cuarta

SIGUIENTE

¿Cuál es tu sexo?

- Hombre
- Mujer

¿Con cuál genero te identificas?

- Transmasculino
- Transfemenino
- Masculino
- Femenino
- No binario
- Otro (por favor especifique)
  [______________]
- No sé
- Prefiero no responder

¿Actualmente te identificas como...?

- Gay (atracción de un hombre hacia otro hombre)
- Lesbiana (atracción de una mujer hacia otra mujer)
- Bisexual (atracción hacia más de un sexo o género)
- Heterosexual (atracción hacia el sexo opuesto)
- Utilizo un término diferente (por favor especifique)
  [____________]
- No sé
- Prefiero no responder

SIGUIENTE

¿Cuál es tu religión o credo?

- Católica
- Evangélica o protestante
- Judía
- Musulmana
- Mormón
- Católica Ortodoxa
- Budista
- Hinduista
- Fe Bahá'í
- Testigo de Jehová
- Otra religión o credo (por favor especifique)
  [____________]
- Ninguna
- Prefiero no responder

En cuestiones políticas, la gente habla de "la izquierda" y "la derecha". ¿En qué punto de esta escala, donde el 1 es izquierda y el 10 es derecha, te ubicarías?

| 1 | 2 | 3 | 4 | 5 | 6 | 7 | 8 | 9 | 10 | No sé / Prefiero no responder |

SIGUIENTE

Queremos asegurarnos de que estás leyendo estas preguntas y no tomando decisiones al azar. Por lo tanto, selecciona la última opción para esta pregunta.

- Primera
- Segunda
- Última

SIGUIENTE

¿En cuál de los siguientes rangos se encuentran los ingresos familiares mensuales de su hogar, incluyendo las remesas del exterior, programas de ayuda en dinero del gobierno o municipio, pensiones o jubilaciones, rentas y el sueldo o ingreso de todos los adultos e hijos que viven en su hogar?

- Entre $0 y $100.000 pesos
- Entre $100.001 y $200.000 pesos
- Entre $200.001 y $350.000 pesos
- Entre $350.001 y $475.000 pesos
- Entre $475.001 y $600.000 pesos
- Entre $600.001 y $700.000 pesos
- Entre $700.001 y $815.000 pesos
- Entre $815.001 y $975.000 pesos
- Entre $975.001 y $1.200.000 pesos
- Entre $1.200.001 y $1.450.000 pesos
- Entre $1.450.001 y $1.600.000 pesos
- Entre $1.600.001 y $1.800.000 pesos
- Entre $1.800.001 y $2.000.000 pesos
- Entre $2.000.001 y $2.400.000 pesos
- Más de $2.400.000 pesos
- No sé / No entiendo la pregunta
- Prefiero no responder

En esta parte de nuestra encuesta, queremos saber qué piensas sobre las percepciones públicas en ciertos temas en Chile. Al responder las siguientes preguntas, por favor piensa en la población **chilena adulta en general.** Más abajo, por favor, mueva el cursor hasta el número que usted considere para cada afirmación.

"En la población chilena adulta, creo que aproximadamente _____ de cada 100 personas se sentirían cómodas...

| 0 | 10 | 20 | 30 | 40 | 50 | 60 | 70 | 80 | 90 | 100 |

...supervisando a un empleado gay."

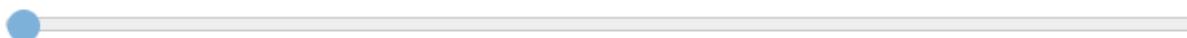

...trabajando estrechamente con un compañero de trabajo gay."

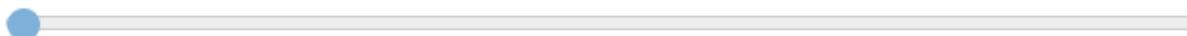

...siendo atendido en el supermercado por un cajero que es gay."

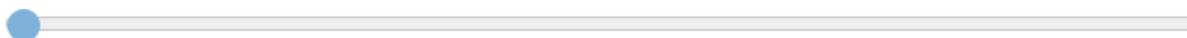

SIGUIENTE

Gracias por sus respuestas. Está cerca de finalizar la encuesta.

Al final del estudio, haremos un sorteo por un monto adicional que usted podrá elegir donarlo a la fundación "**Iguales**" o recibirlo usted mismo.

La organización **Iguales** trabaja "para conseguir la plena inclusión de la diversidad sexual en la sociedad chilena. Para ello, participan en todas las etapas de la formulación de políticas públicas a nivel legislativo y administrativo."

50 participantes serán elegidos al azar para recibir $100.000 pesos. Si usted fuera elegido, recibirá o donaremos el dinero de acuerdo a su respuesta y se le notificará directamente a su correo para hacer efectivo su premio.

**Todos los participantes tendrán la misma probabilidad de ser elegidos, sin importar sus respuestas.**

**Su respuesta no afectará la probabilidad de ganar el sorteo.**

¿Qué porción de estos $100.000 pesos adicionales preferirías donar a la fundación **Iguales**?

*Por favor, notar que los investigadores no tienen **una afiliación ni conflicto de interés** con la fundación Iguales. La elección de incluirla en este estudio fue recomendada por una tercera parte independiente.*

0    10000    20000    30000    40000    50000    60000    70000    80000    90000    100000

**Monto a donar**

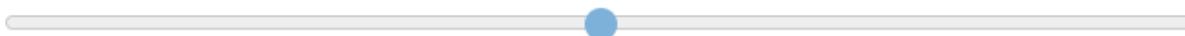

SIGUIENTE

Por favor, responda las siguientes preguntas sobre nuestro estudio.

Indique en qué medida está de acuerdo o en desacuerdo con la siguiente afirmación:

**"Las instrucciones fueron claras"**

- Totalmente de acuerdo
- Algo de acuerdo
- Ni de acuerdo ni en desacuerdo
- Algo en desacuerdo
- Totalmente en desacuerdo

¿Hay algo que no esté claro o que sea confuso en el estudio?

¿Hay algo más que le gustaría compartir con los investigadores?

SIGUIENTE

Gracias por sus respuestas. Lamentablemente, usted no ha sido elegido para recibir un bonus adicional de $100.000 pesos.

**Finalmente, por favor haga clic en el siguiente botón para finalizar. No cierre esta ventana antes de llegar al final.**

[SIGUIENTE]

Gracias por dedicarle tiempo a esta encuesta.
Su respuesta se ha registrado.

# References used in the Online Appendix